\documentclass[]{aa}
\usepackage{graphicx}
\usepackage{media9}
\usepackage[varg]{txfonts}
\usepackage{url}
\newcommand{\kms}{km\,s$^{-1}$}
\usepackage{natbib}
\bibpunct{(}{)}{;}{a}{}{,}
\usepackage[dvipsnames]{xcolor}
\usepackage[normalem]{ulem}

\usepackage{color}

\begin{document} 
\title{V838 Monocerotis as seen by ALMA:\\ a remnant of a binary merger in a triple system}
   \author{Tomek Kami\'nski\inst{1} 
          \and Romuald Tylenda\inst{1}
          \and Aleksandra Kiljan\inst{2}
          \and Mirek Schmidt\inst{1}
          \and Krzysztof Lisiecki\inst{3}
          \and Carl Melis\inst{4}
          \and Adam Frankowski\inst{1}
          \and Vishal Joshi\inst{5}
          \and Karl M.\ Menten\inst{6}
          }
\institute{\centering
Nicolaus Copernicus Astronomical Center, Polish Academy of Sciences, Rabia{\'n}ska 8, 87-100 Toru{\'n}, \email{tomkam@ncac.torun.pl}\label{inst1}
\and Warsaw University Astronomical Observatory, Al. Ujazdowskie 4, 00-478 Warszawa, Poland \label{inst2}
\and Institute of Astronomy, Nicolaus Copernicus University in Toru{\'n}, Grudzi\k{a}dzka 5, 87-100 Toru{\'n}, Poland\label{inst3}
\and Center for Astrophysics and Space Sciences, University of California, San Diego, CA 92093-0424, USA\label{inst4}
\and Physical Research Laboratory, Navrangpura, Ahmedabad, Gujarat 380009, India\label{inst6}
\and Max-Planck-Institut f\"ur Radioastronomie, Auf dem H\"ugel 69, 53-121 Bonn, Germany \label{inst7}
}
\authorrunning{T. Kami\'nski et al.}
\abstract{V838\,Mon erupted in 2002 quickly becoming the prototype of a new type of stellar eruptions known today as (luminous) red novae. The red nova outbursts are thought to be caused by stellar mergers. The merger in V838\,Mon took place in a triple or higher system involving two B-type stars.}
{We  wish to characterize the merger remnant $\sim$17\,yr after the eruption to learn more about the remaining system, the progenitor, and merger physics.}
{We mapped the merger site with ALMA at a resolution of $\sim$25\,mas, or 148\,au for a distance of 5.9\,kpc, in continuum dust emission and in rotational lines of simple molecules, including CO, SiO, SO, SO$_2$, AlOH, and H$_2$S. We use radiative transfer calculations to reproduce the remnant's architecture at the epoch of the ALMA observations.}
{For the first time, we identify the position of the B-type companion relative to the outbursting component of V838\,Mon. The stellar remnant is surrounded by a clumpy wind with characteristics similar to winds of red supergiants. The merger product is also associated with an elongated structure, $17.6 \times 7.6$\,mas ($104\times 45$\,au), seen in continuum emission, and which we interpret as a disk seen at a moderate inclination. Maps of continuum and molecular emission show also a complex region of interaction between the B-type star (its gravity, radiation, and wind) and the flow of matter ejected in 2002. The remnant's molecular mass is about 0.1\,M$_{\sun}$ and the dust mass is 8.3$\cdot$10$^{-3}$\,M$_{\sun}$. The mass of the atomic component remains unconstrained.}
{The most interesting region for understanding the merger of V838\,Mon remains unresolved but appears elongated. To study it further in more detail will require even higher angular resolutions. ALMA maps show us an extreme form of interaction between the merger ejecta with a distant ($\sim$250\,au) companion. This interaction is similar to that known from the Anteres AB system but at a much higher mass loss rate. The B-type star not only deflects the merger ejecta but also changes its chemical composition with an involvement of circumstellar shocks. The ALMA view on V838\,Mon offers the best so far images of a merger site.} 

\maketitle

\section{Introduction}\label{intro}
V838 Monocerotis (V838\,Mon) underwent a red-nova event in early 2002 \citep[][and references therein]{Munari2002,kimeswenger,Crause2003,TylendaProgenitor,TylendaEvolutionV838}. Its spectacular outburst to a luminosity of 10$^6$\,L$_{\sun}$ \citep{sparks,TylendaEvolutionV838} was associated with an iconic light echo \citep{BondEcho,TylendaEcho,TylendaKaminskiEcho}, reflected in a near interstellar cloud \citep{KamiEcho}, and which drew a lot of attention to V838\,Mon, even among non-astronomers. The leading hypothesis explaining its eruption is a merger between two young \citep[$<$25\,Myr;][]{KamiSecond,AfsarBond} stars \citep{TS2006} of approximate masses of 8.0 and 0.4\,M$_\sun$ \citep{TylendaProgenitor}. Spectra obtained during the eruption showed that the 2002 event produced outflows with velocities approaching 500\,\kms\ \citep[e.g.,][]{WisniewskiSpecpol}. After three months of high but varying brightness, the object entered a phase of decline which, characteristically for red novae, resulted in a very cool remnant with a photospheric temperature possibly as low as 2000--2300\,K \citep{evans,pavlenko}. Since then, the circumstellar material became primarily apparent through atomic lines and molecular bands characterized by a very low excitation, typically of a few hundred K \citep{KamiKeck}. In the cool post-outburst phase, in late 2002, the merger ejecta started forming dust \citep{TylendaEvolutionV838,WisniewskiPhoto}. In this paper, we present observations of the cool circumstellar remnant about 17\,yr after the eruption. With submillimeter observations at 20\,mas resolution, for the first time we spatially resolve both gas and dust in the circumstellar environment. 

In early observations of the outburst, it was recognized that the erupting object lies close to a B-type main-sequence star \citep{Bdiscovery,Bdiscovery2}. Follow-up observations revealed that it is a physical companion to V838\,Mon \citep{TylendaEngulf} and that with a few other B-type stars in its surrounding it belongs to a scant open cluster \citep{AfsarBond,Ortiz-Leon} in the far outer Galaxy, at a distance of 5.9\,kpc \citep{sparks,Ortiz-Leon}. Pre-outburst photometry suggests that V838\,Mon was an early B type star as well \citep{TylendaProgenitor}. The presence of the companion makes the merger scenario of V838\,Mon particularly interesting as it happened in a triple (or a higher multiple) system. 

Not much is known about the B-type companion. A few of its spectral features, always entangled with the much brighter spectrum of the luminous red remnant of the merger, revealed its spectral type B3\,V (or effective temperature of 18\,000\,K) and a fast rotation rate of $v \sin i$=250$\pm$50\,\kms\ \citep{KamiKeck}. It however appears to be under-luminous compared to the other B type stars in the cluster, a problem which has not been explained \citep{TylendaKaminskiEcho}. In 2005, the signatures of the B star disappeared. This was first interpreted as an eclipse \citep{eclipse1,eclipse2}, but further studies showed that the blue star vanished because it became completely embedded in the dusty ejecta produced in the 2002 eruption \citep{TylendaEngulf}. Prior to the complete disappearance from the optical, the B star started photo-ionizing the approaching ejecta producing an emission spectrum that included  relatively strong lines of [\ion{Fe}{II}] \citep{eclipse2,KamiKeck}. Based on these features and the timing of the ``eclipse'', \citet{TylendaEngulf} found that the separation of V838\,Mon and the B star is about 250\,au. That calculation relied on the gas velocity which was not straightforward to infer, given that the outflow velocities varied considerably over the 2002 event, and it was uncertain whether spherical symmetry can be assumed for the merger remnant. Here we present observations which for the first time reveal directly the location of the B star with respect to V838\,Mon. 

Currently, the stellar remnant of V838\,Mon is a luminous and cool star with a wind-like outflow \citep{TylendaEngulf} and active but decaying SiO masers \citep{Deguchi2005,Ortiz-Leon}, which are commonly associated with outflows from mass-losing asymptotic giant branch (AGB) stars and red supergiants. V838\,Mon shares a lot more of its observational characteristics with red supergiants. As an M type ''supergiant'' with a main-sequence B-type companion, the system can be technically classified as of the rare VV\,Ceph type \citep{binarySupergiants,VVCeph}. Here we demonstrate more specifically that it can be treated as an analog of the binary system of Antares but with a much denser outflow and much more pronounced interaction effects between the cool dusty wind and the hot star.

This paper is organized as follows. In Sect.\,\ref{sec-obs}, we present technical details on the high-resolution observations of V838\,Mon with the Atacama Large Millimeter/submillimeter Array (ALMA). The following two sections present results of these observations in continuum (Sect.\,\ref{sec-res-conti}) and in line emission (Sect.\,\ref{sec-res-lines}). Section\,\ref{sec-optical} briefly introduces optical observations of the remnant before and after the ALMA epoch to indicate the status of V838\,Mon in the context of the temporal evolution since the end of the outburst. In Sect.\,\ref{sec-3d} we present our three-dimensional (3D) reconstruction of the remnant, primarily based on the dust emission data. We discuss all results in Sect.\,\ref{sec-discussion} and summarize them in Sect.\,\ref{sec-summary}.

\section{ALMA observations}\label{sec-obs}
The ALMA observations were acquired on 20 June 2019 in three consecutive execution blocks. The array with 43 antennas was in its most extended configuration with baselines of 83.1--15\,238.4\,m. Visibilities were calibrated with default calibration scripts delivered by the observatory and data were processed, including imaging, with the Common Astronomy Software Applications \citep[CASA;][]{casa}. An attempt to improve data quality by a self-calibration procedure was unsatisfactory and visibilities with the original calibration were used. The observations were executed in band 6 and covered frequencies 215.56--217.44, 219.06--220.94, 230.11--231.99, and 232.06--233.94 GHz at a native resolution of 1.3--2.5\,\kms. These spectral ranges overlap with older interferometric observations of V838\,Mon with the Submillimeter Array (SMA) which however have a lower angular resolution and a much worse sensitivity \citep{KamiSubmm}. The continuum sensitivity of the ALMA data is 9.8\,$\mu$Jy per beam at a beam full width at half-maximum (FWHM) of 28$\times$25 mas and at natural weighting of visibilities. Continuum images include only line-free channels and cubes with line emission had continuum removed by a polynomial fit to calibrated visibilities. Parts of the ALMA dataset has been briefly presented in \citet{Ortiz-Leon}.

\section{ALMA results: continuum}\label{sec-res-conti}
\begin{figure*}
    \centering
    \includegraphics[width=0.45\textwidth]{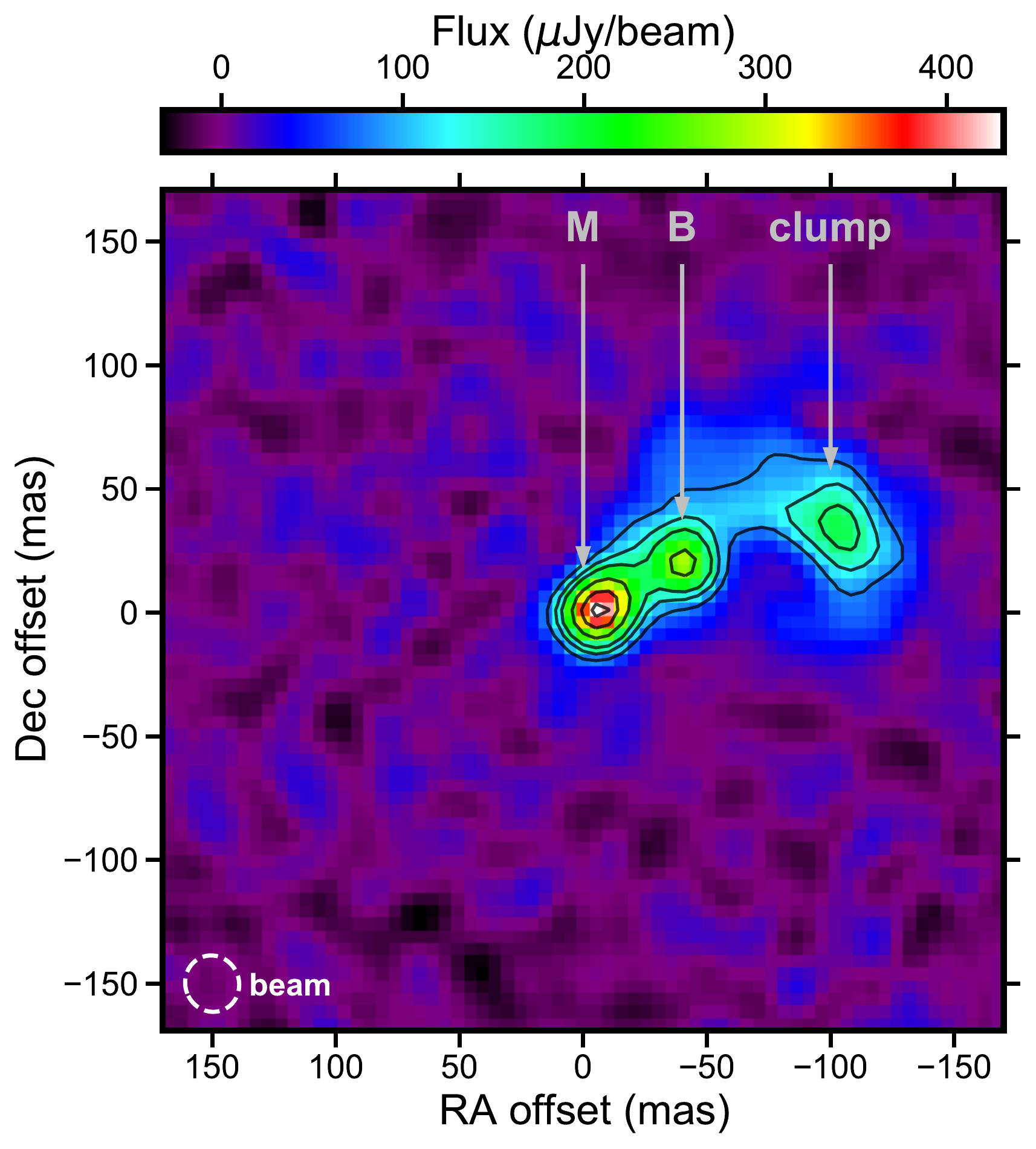}
    \includegraphics[width=0.45\textwidth]{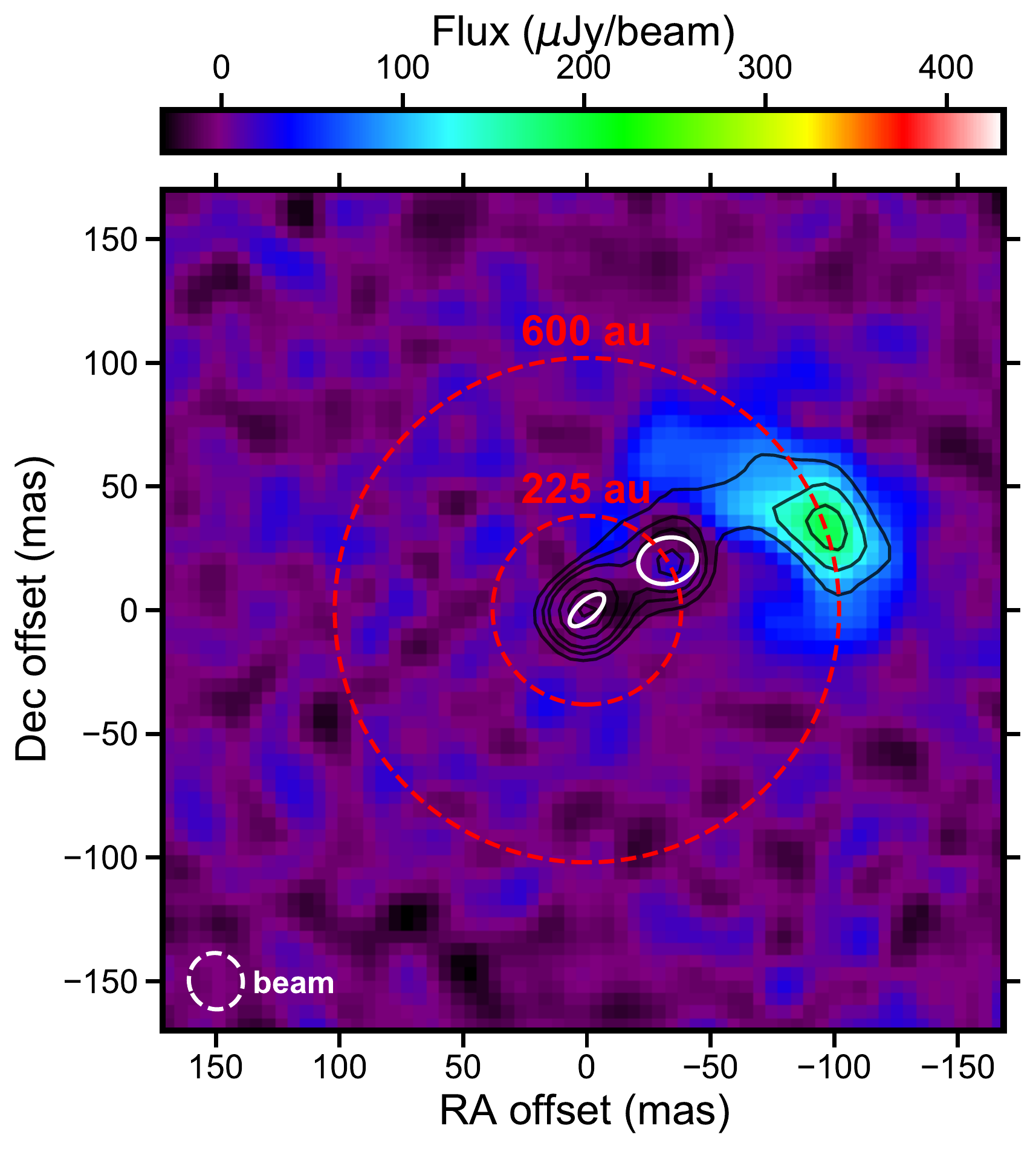}
    \caption{Millimeter-wave emission of V838\,Mon. Left: ALMA map at 1.3 mm with all discussed sources labeled. Right: Residual emission after subtracting Gaussian fits to components M and B. Ellipses drawn with full line show the FWHM of the fitted Gaussian components. White ellipses drawn with dashed line show FWHM of the restoring beam. Contours in both images show the original continuum emission at 0.2, 0.3, 0.4, 0.6, 0.8, and 0.95 times the peak emission. The maps have the same colorscale. The spatial offsets are with respect to the best-fit central position of the M component.}
    \label{fig-conti-residuals}
\end{figure*}

We first present results for the continuum emission imaged at an effective frequency of 224.75\,GHz or 1.334\,mm. For the analysis, images were produced at different weighting schemes. An example map with Briggs weighting and the robust parameter of $R$=0.5 (resulting in a beam of a FWHM of 22\,mas) is shown in Fig.\,\ref{fig-conti-residuals}. There are three main emission components seen in the continuum maps. The strongest one is centered at ICRS coordinates RA=07:04:04.821535($\pm$0.000012) and Dec=$-03$:50:50.629426($\pm$0.000177) (J2000.0).  The location is consistent with a VLBI position of V838\,Mon's masers \citep{Ortiz-Leon}. The continuum source can be identified as emission of circumstellar material in the direct vicinity of the stellar remnant of the 2002 eruption. We call it hereafter ``component M''. A Gaussian fit to this compact source (at Briggs' $R$=0.5) gives a size (corrected for the beam-size) of (17.6$\pm$0.8)$\times$(7.6$\pm$1.4)\,mas with the major axis at a position angle (PA) of 132\degr$\pm$4\degr. Its flux of 588$\pm$14\,$\mu$Jy constitutes 28\% of the total continuum flux in the map. The position angle of the submm component is close to PA=170\degr$\pm$30\degr\ of an elongated MIR structure surrounding V838\,Mon and found in VLTI observations by \citet{Olivier} in 2011 and by Mobeen at al. (in prep.) in 2019. 

The secondary component is centered at ICRS coordinates RA=07:04:04.819359($\pm$0.000051) and Dec=--03:50:50.609454($\pm$0.000965), or 32.6\,mas west and 20.0\,mas north (38.2\,mas or 225.4\,au at a PA of --58\fdg5) from the main component. A Gaussian fit yields a beam-deconvolved size of (24.0$\pm$3.5)$\times$(18.5$\pm$3.5) mas at a PA of 166\degr$\pm$45\degr\  (i.e. is nearly circular) and a flux of 510$\pm$48 $\mu$Jy (24.4\% of the total flux in the map). We identify the component as material directly surrounding the B-type companion of V838\,Mon and call it hereafter the ``B component''. Its projected distance to V838\,Mon of 225\,au (at the 5.9\,kpc distance) is close to the separation $a\approx$250\,au calculated from  the timing of the interaction of the B-type star with the merger ejecta \citep{TylendaEngulf}. This close match between the physical and projected distances suggests further that the binary orbit is almost in the sky plane. There is no submillimeter emission bridging the two stars, that is, the two Gaussian components account for all emission seen around the stars at our sensitivity level.

The third component is much more extended and contains all the remaining emission (47.6\%  of the total or 1.0\,mJy). It has an irregular shape and an approximate size of 135$\times$70 mas. Its outer parts have low surface brightness and maps at a better sensitivity may reveal that it is even more extended. Hereafter, we refer to it as ``the external cloud'' or ``the clump''. Its peak emission is located at $\approx$100\,mas or 590\,au from the V838\,Mon position. A line connecting the peak cloud emission with that of V838\,Mon does not cross directly the position of the B component, but there are cloud parts directly at the line connecting M and B (see the residual map in Fig.\,\ref{fig-conti-residuals}). 
Some weak diffuse emission is seen north and south-west of and directly next to the B component, so there may be some physical continuation between the cloud and the B component. 

\section{ALMA results: spectral lines}\label{sec-res-lines}
\begin{figure*}
    \centering
    \includegraphics[width=0.99\textwidth, trim=0 0 0 120]{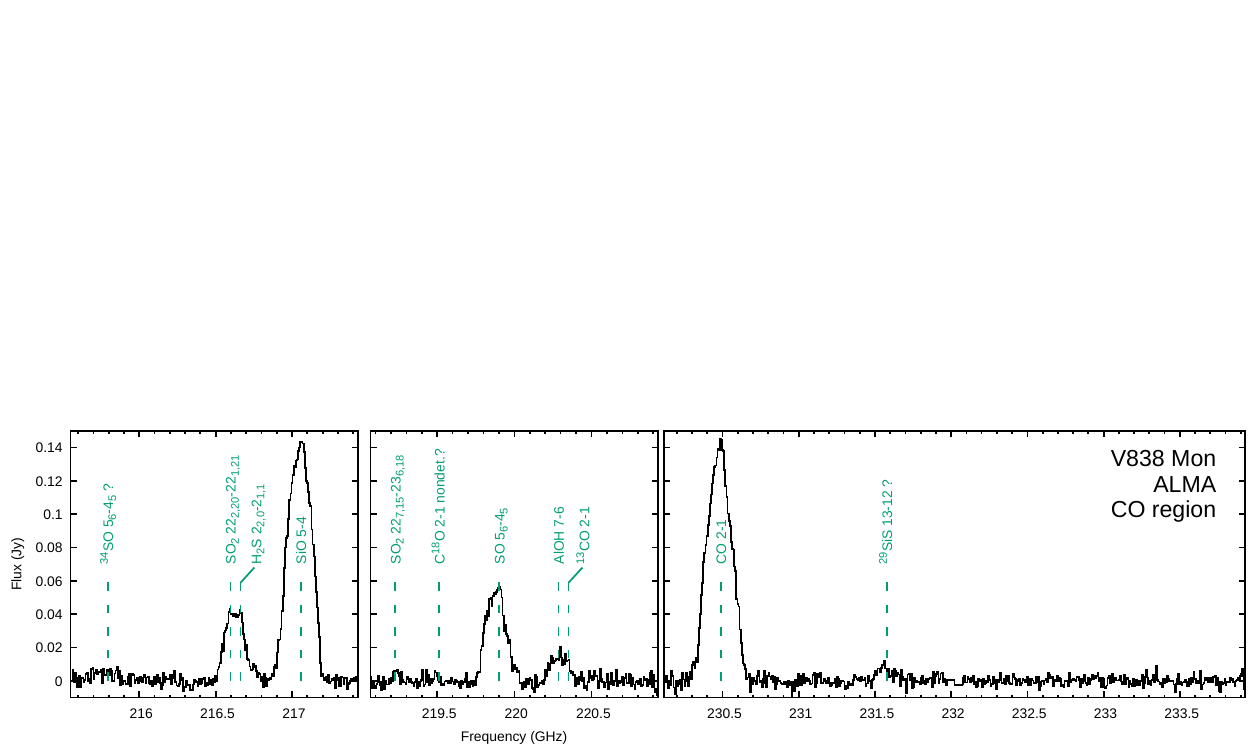}
    \caption{ALMA spectrum of V838\,Mon. It represents the entire emission region of CO. Main identified features are labeled and marked with dashed lines at frequencies corresponding to the systemic velocity. Uncertain identifications are indicated with a question mark. }
    \label{fig-spec}
\end{figure*}

The observations covered several molecular emission features which are listed in Table\,\ref{tab-lines} and shown in a composite spectrum in Fig.\,\ref{fig-spec}. These include features of common oxides, SiO, CO, SO (along with $^{34}$SO), and SO$_2$, and of H$_2$S. Millimeter-wave emission of these molecules has been known in V838\,Mon since the SMA observations of \citet{KamiSubmm}. We additionally tentatively detected rare isotopologues of $^{13}$CO and $^{29}$SiS. A newly detected species is AlOH which is observed through its $J$=7--6 transition \citep{AlOH} and is seen in the spectral range that was not covered by the earlier SMA observations. It is blended with a flank of the $^{13}$CO 2--1 line, if the latter is really present. 

\begin{table}
    \centering
    \caption{Lines observed in V838\,Mon.}\label{tab-lines}
    \begin{tabular}{llccc}
\hline
Line &\multicolumn{1}{c}{$\nu$}& $E_u$ &$A_{ud}$  &$g_u$\\
     &\multicolumn{1}{c}{(GHz)}& (K)   &(s$^{-1}$)&     \\
\hline       
$^{34}$SO? $5_6-4_5$            &215.8399&  ~34.38 & 1.26e--4 & 13 \\
SO$_2$ $22_{2,20}-22_{1,21}$    &216.6433\tablefootmark{a}&  248.44 & 9.27e--5 & 45 \\
H$_2$S $2_{2,0}-2_{1,1}$        &216.7104\tablefootmark{a}&  ~83.98 & 4.87e--5 & ~5 \\
SiO $5-4$                       &217.1049&  ~31.26 & 5.20e--4 & 11 \\
gap                             &        &         &\\
SO$_2$ $22_{7,15}-23_{6,18}$    &219.276~&  352.76 & 2.13e--5 & 45 \\
SO $5_6-4_5$                    &219.9494&  ~34.98 & 1.34e--4 & 13 \\
AlOH $7-6$                      &220.3344\tablefootmark{a} &  ~42.30 & 6.18e--6 & 12 \\
$^{13}$CO? $2-1$                &220.3987\tablefootmark{a} &  ~15.87 & 6.08e--7 & 10 \\
gap                             &        &         &\\
CO $2-1$                        &230.538~ &  ~16.60 & 6.91e--7 & ~5 \\
$^{29}$SiS? $13-12$             &231.6267 &  ~77.82 & 2.10e--4 & 27 \\
\hline
    \end{tabular}
    \tablefoot{Frequencies ($\nu$), Einstein coefficients for spontaneous emission $A_{ud}$, and statistical weights ($g_u$) are given. Question marks indicate uncertain identification. \tablefoottext{a}{blended}.}
\end{table}

Physical conditions in the molecular gas surrounding V838\,Mon were studied by \citet{KamiSubmm}. There, an LTE excitation analysis was based on SMA observations of multiple lines observed in a much wider spectral range than covered here with ALMA. Since our new ALMA observations provide the actual sizes of the emission regions, we revised the multi-line models by implementing updated beam filling factors. Aside from changes in absolute values of column densities, we obtained very similar results as before. The gas has source-averaged brightness temperatures between around 70 and 200\,K for different molecular tracers. Although ALMA spatially resolved the source, we are unable to investigate temperature gradients through line ratios because for each species ALMA covered only a single transition. Our radiative transfer calculations show, however, that the emission of CO 2--1 is moderately optically thick ($\tau \approx 1$) at line cores at positions with strong emission. In line wings and most other positions the emission is optically thin. Brightness temperatures (per synthesized beam) of CO 2--1 are of 30--170\,K. Assuming a beam filling factor of one everywhere and given the moderate optical depth of the emission, this range is a rough measure of excitation temperatures at the CO photosphere. We find column densities per beam of $10^{20}$\,cm$^{-2}$ or lower. These values are slightly higher than in \citet{KamiSubmm}, but they yield an overall mass of H$_2$ of 0.1\,M$_{\sun}$ that is consistent with the old data.    

The ALMA data do not warrant good measurements of the ratios of column densities for CO isotopologues, even though $^{12}$CO, $^{13}$CO, and C$^{18}$O were covered (cf. Fig.\,\ref{fig-spec}). Nevertheless, by simulating the spectra under LTE, we find a line ratio $^{12}$CO/$^{13}$CO to be larger than about 40. This ratio is hard to constrain because the weaker $^{13}$CO line blends with the wider and much stronger AlOH 7--6 line. The C$^{18}$O is not detected, yielding only an upper limit on the CO/C$^{18}$O ratio of about 80. These limits are consistent with earlier, but equally uncertain, constraints of  $^{12}$CO/$^{13}$CO$\approx$10--100 \citep{Geballe}. 

\begin{figure*}
    \centering
    \includegraphics[width=0.33\textwidth, trim=0 0 0 0]{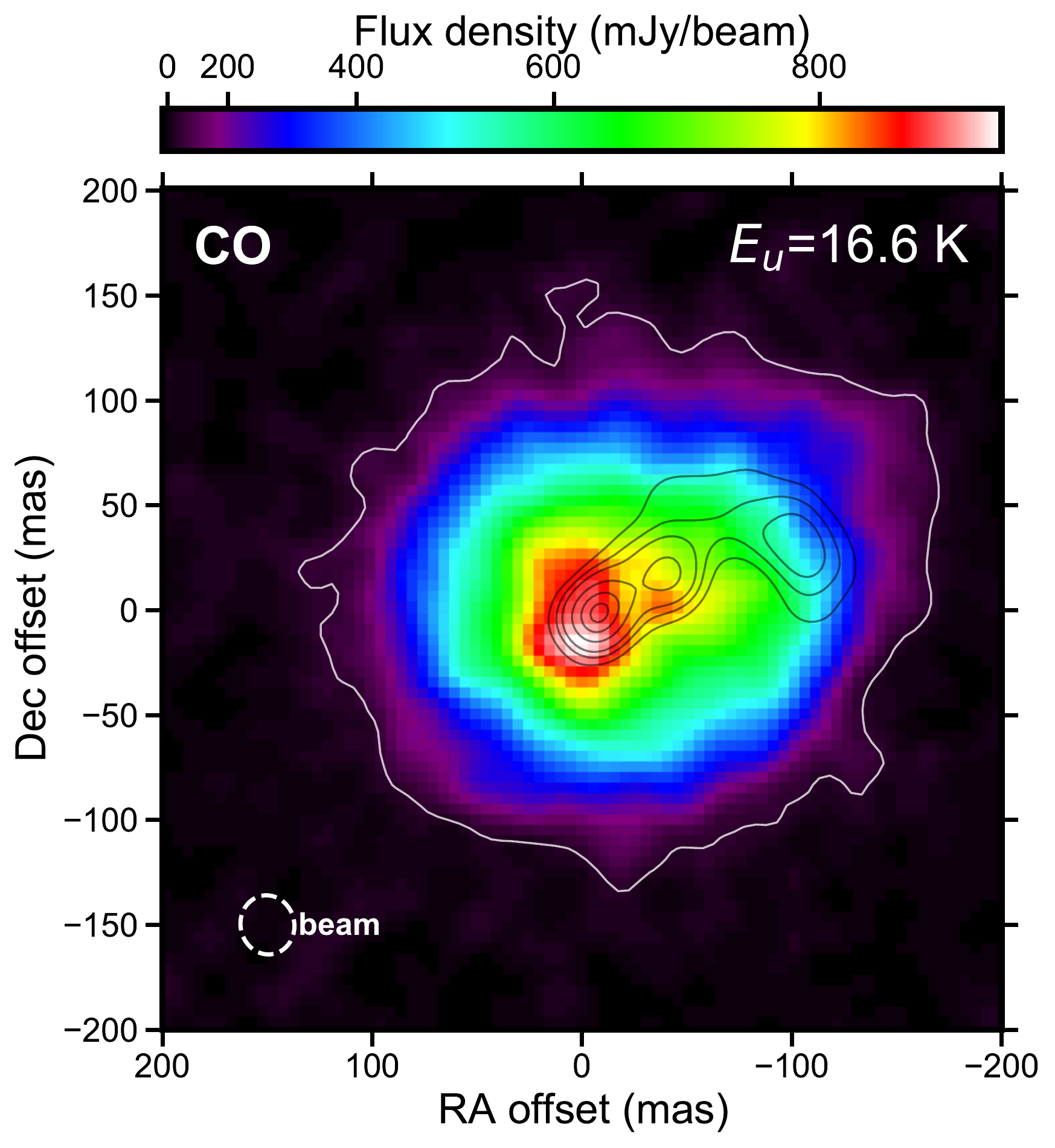}
    \includegraphics[width=0.33\textwidth, trim=0 0 0 0]{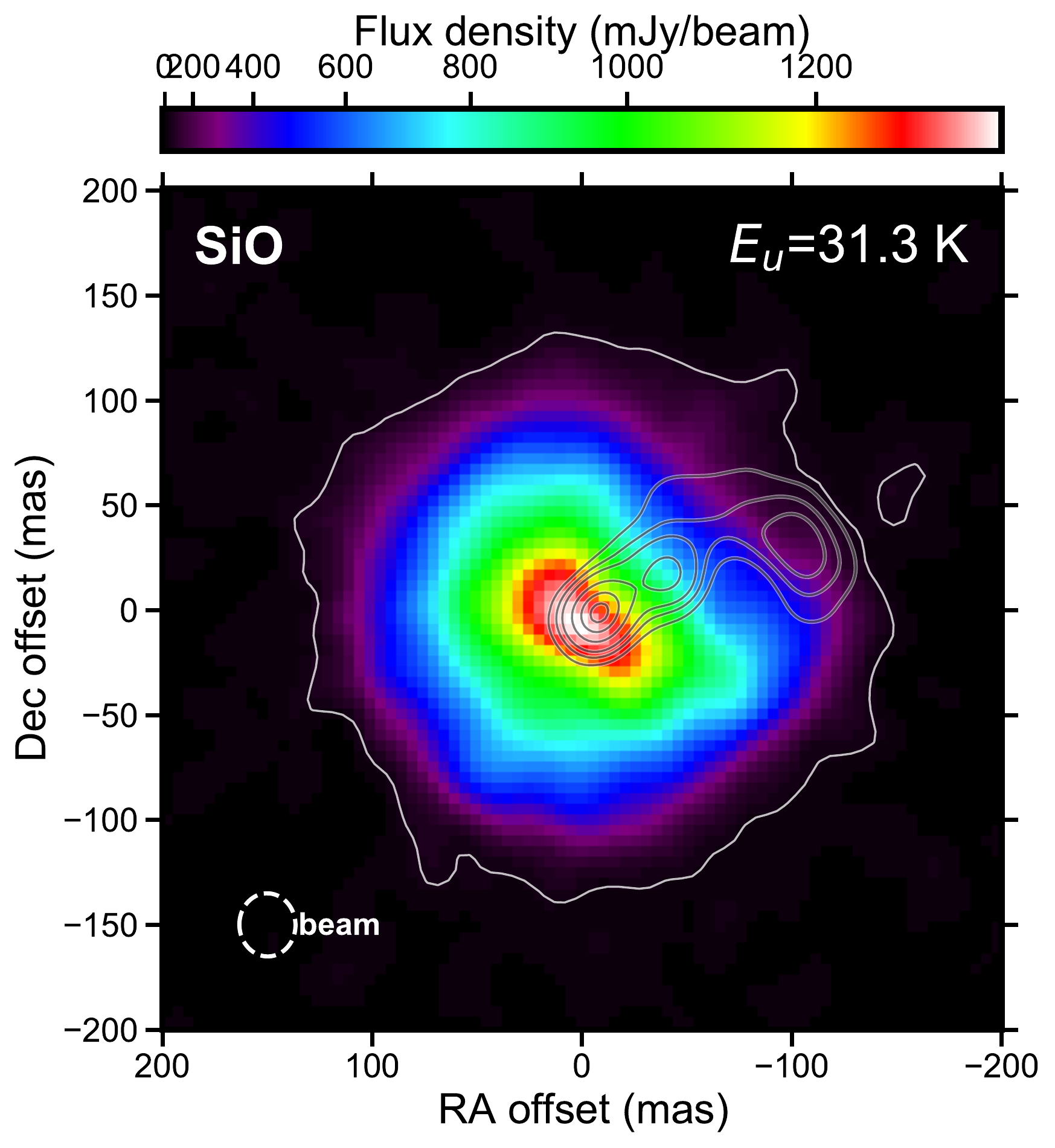}
    \includegraphics[width=0.33\textwidth, trim=0 0 0 0]{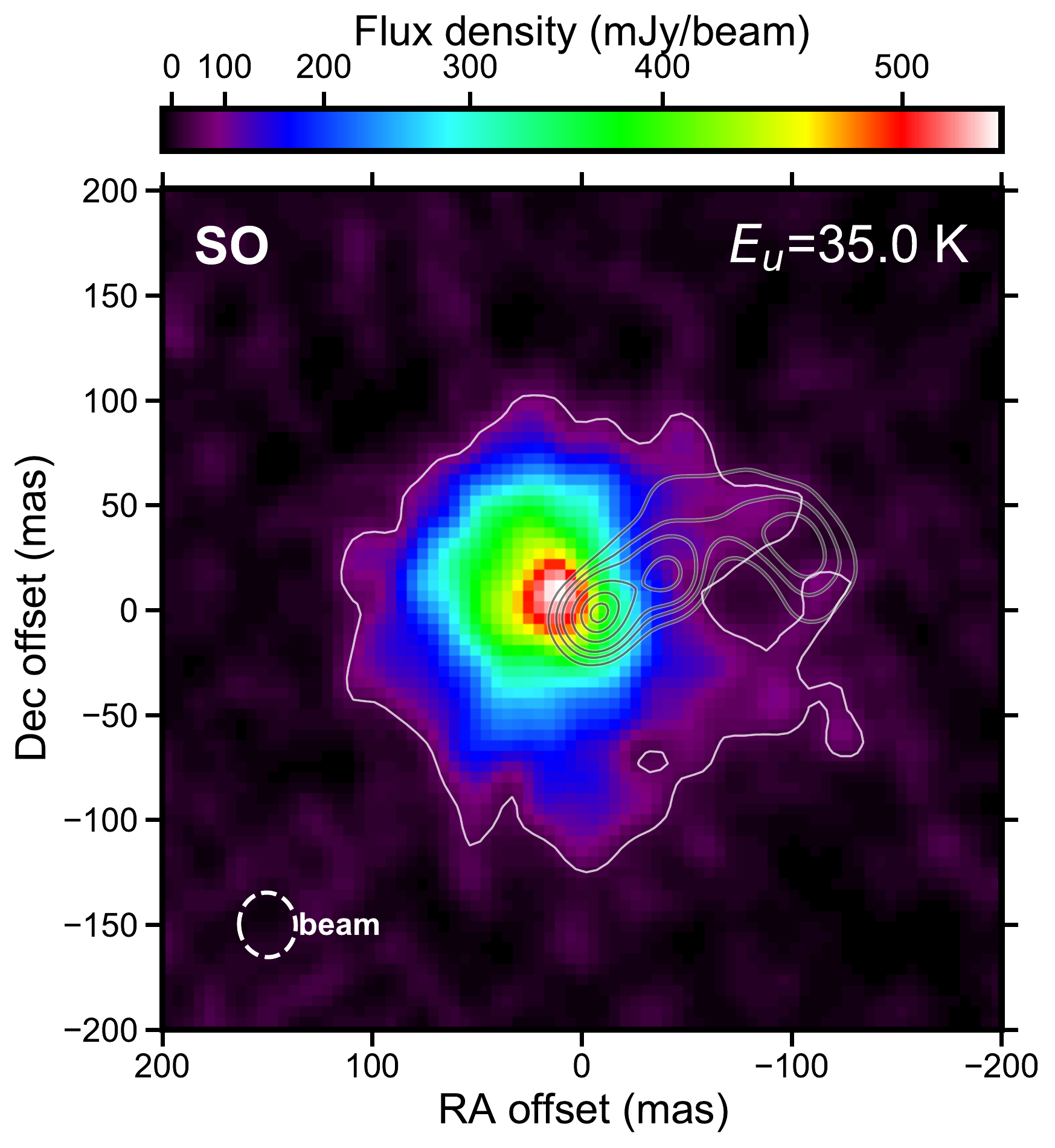}
    \includegraphics[width=0.33\textwidth, trim=0 0 0 0]{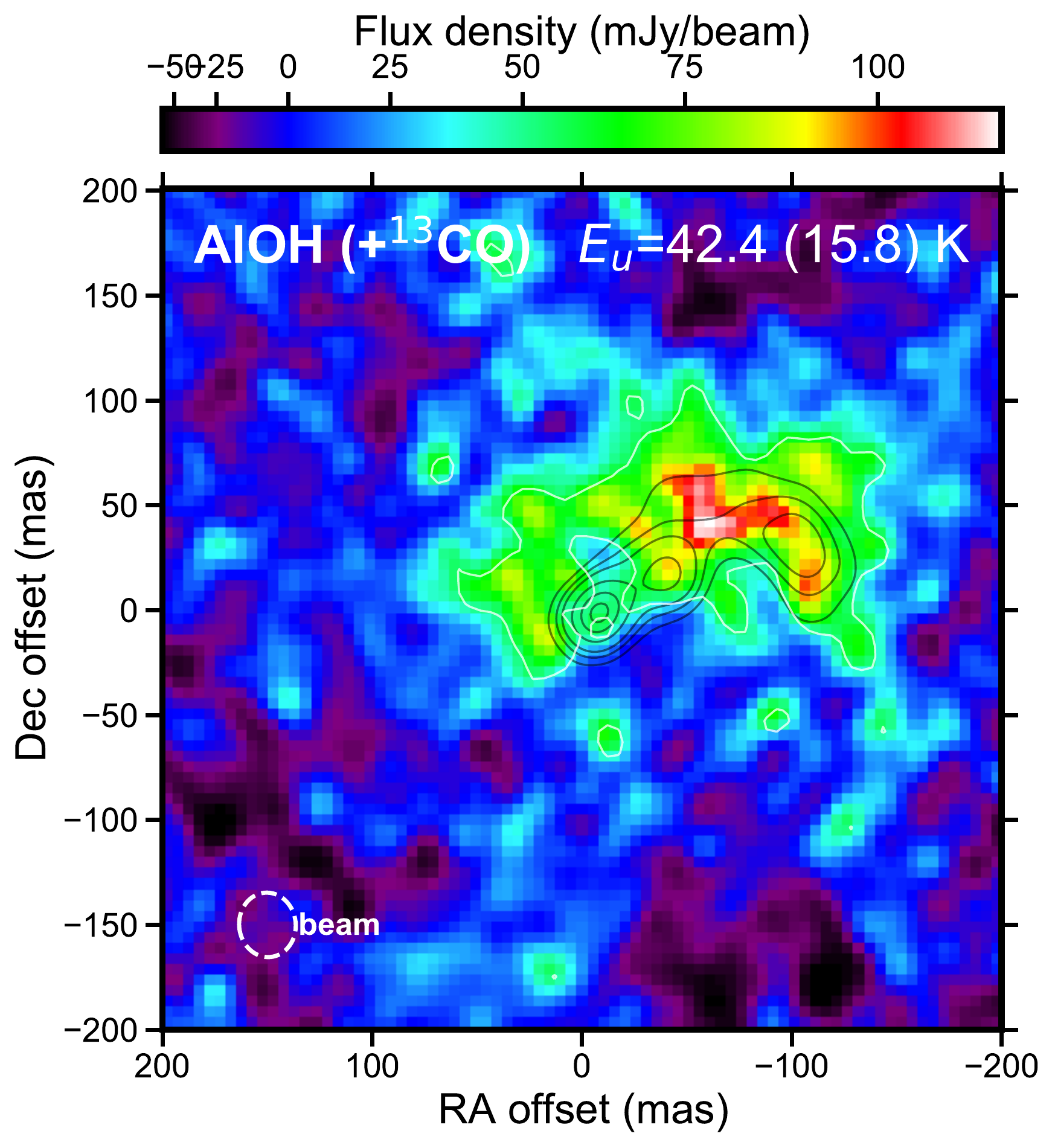}
    \includegraphics[width=0.33\textwidth, trim=0 0 0 0]{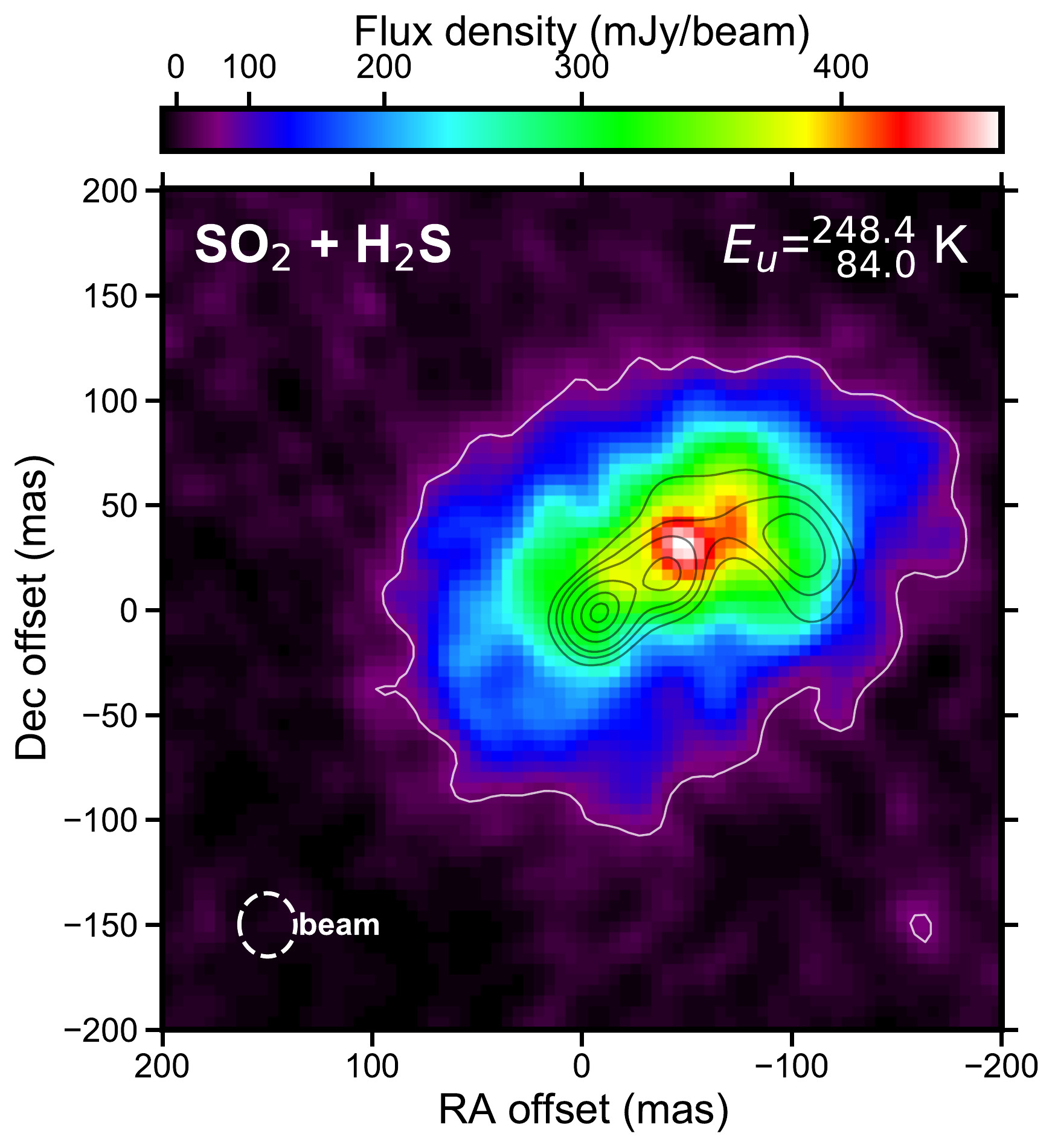}
    \includegraphics[width=0.33\textwidth, trim=0 0 0 0]{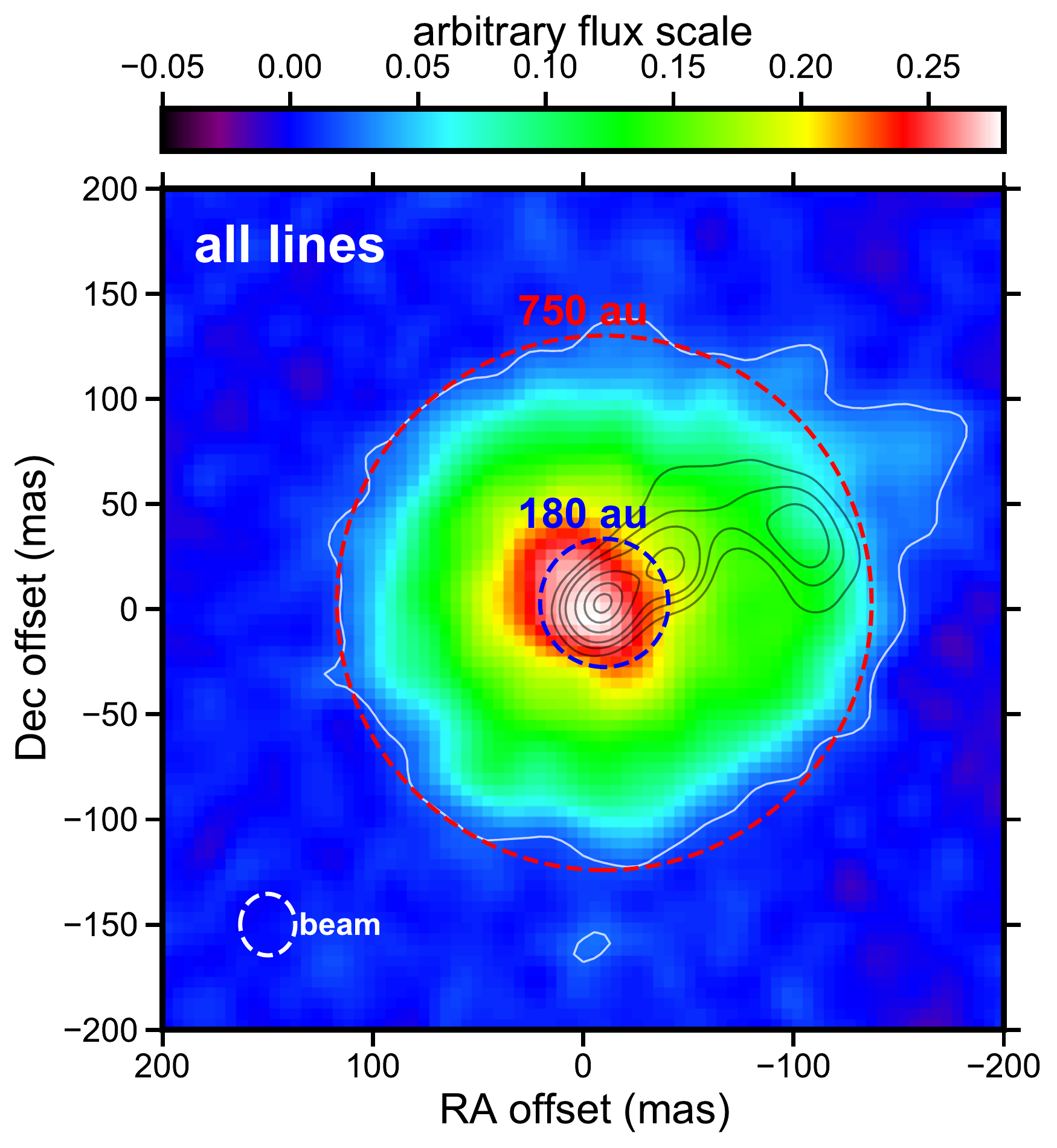}
        \caption{Total intensity maps of the molecular emission in V838\,Mon. White contour is drawn at the 3$\sigma$ noise level of each map. Grey contours show continuum emission at 0.2, 0.3, 0.4, 0.6, 0.8, and 0.95 times the peak level. All maps were reconstructed at natural weighting. The bottom left and middle panels show maps for blended transitions. The energy of the upper level of the mapped transitions is indicated on each map. Bottom right panel represents emission in all lines covered by ALMA. The red dashed circle of a radius of 750\,au serves as the reference for the physical scale.}
    \label{fig-line-maps}
\end{figure*}

Interferometric maps obtained with ALMA reveal detailed distributions of emission for five strongest spectral features, three of which, namely CO (2--1), SiO (5--4), and SO ($5_6$--$4_5$), are transitions that are relatively uncontaminated by emission of other species. We use the three lines to trace the spatio-kinematical distribution of the gas. The upper energy levels for the transitions are, respectively, 16.6, 31.3, and 35.0\,K. The maps of velocity-integrated emission (zeroth moment) are shown in Fig.\,\ref{fig-line-maps}. The distribution of gas emission is different from that of the dust continuum emission. 

\begin{figure*}
    \sidecaption
    \includegraphics[width=0.33\textwidth, trim=0 0 0 0]{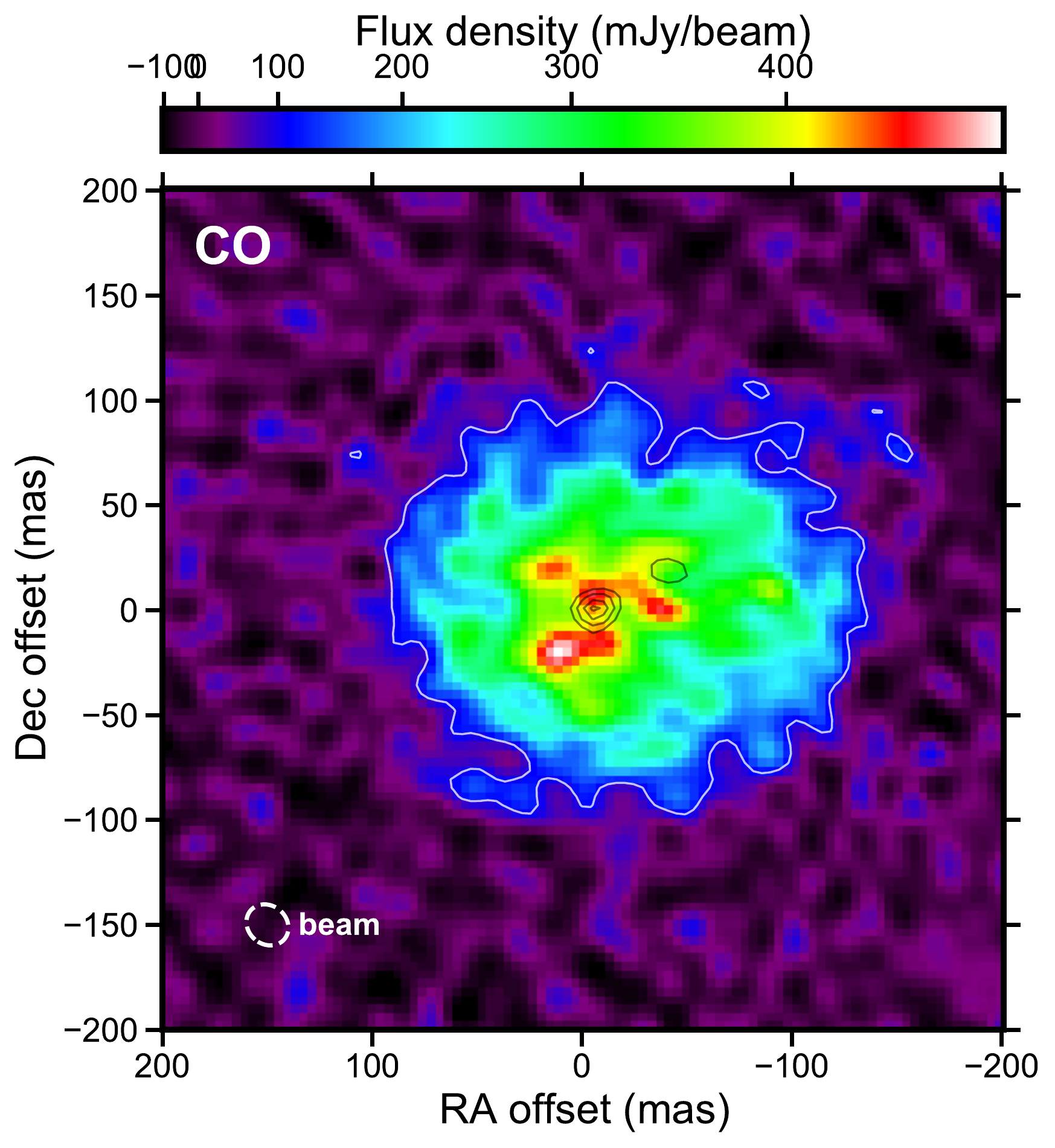}
    \includegraphics[width=0.33\textwidth, trim=0 0 0 0]{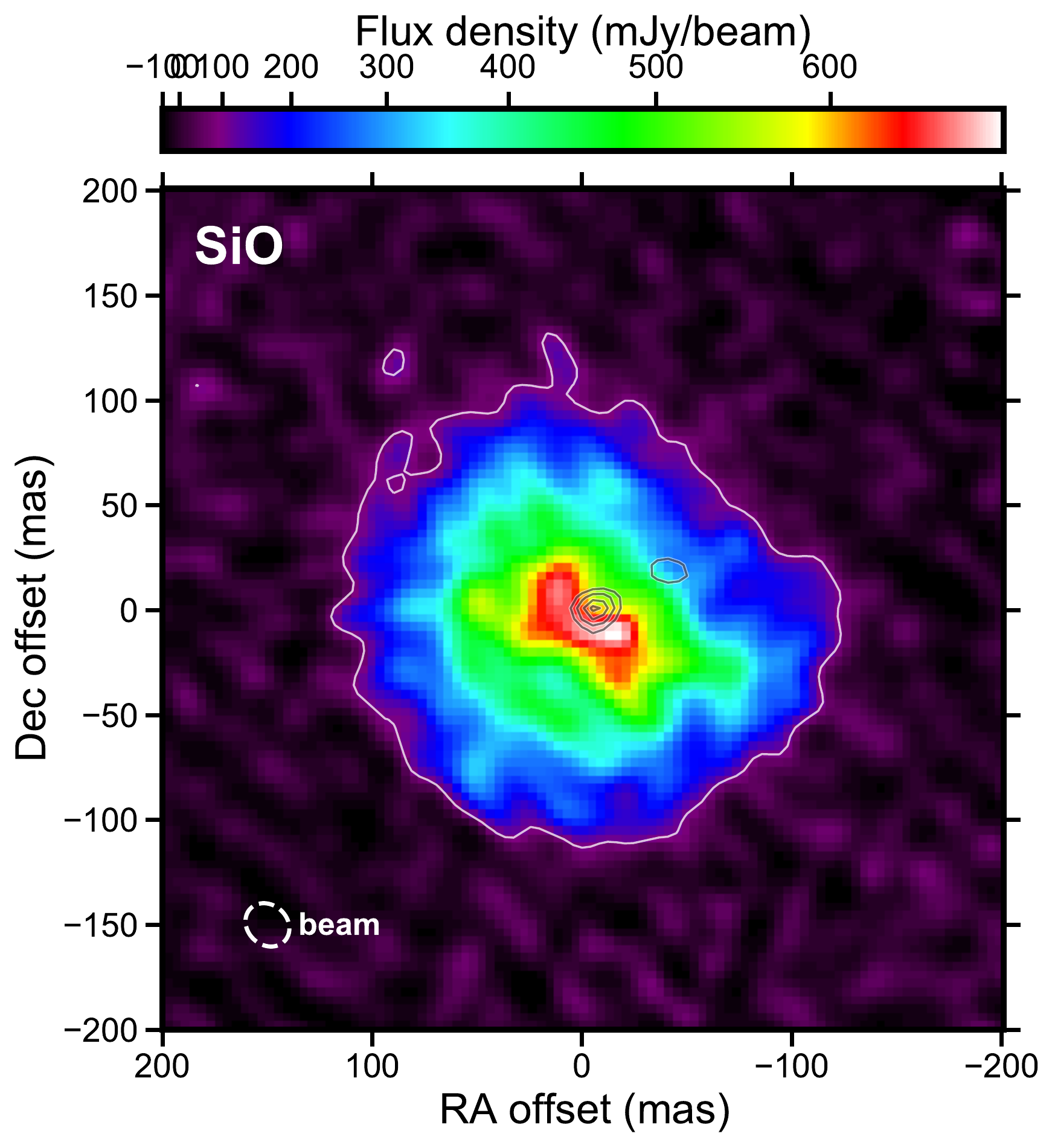}
    \caption{The same as in Fig.\,\ref{fig-line-maps} but for CO and SiO at uniform weighting of visibilities. Continuum contours, also at uniform weighting, start at 0.3 times peak level.}
    \label{fig-line-maps-uni}
\end{figure*}

We assume that CO is the most representative probe of the entire molecular remnant, that is, it can be used as a proxy of the H$_2$ distribution. The CO emission is the most extended and fairly symmetric (Fig.\,\ref{fig-line-maps}). On maps reconstructed with natural weighting the 5$\sigma$ contour extends up to a radius of 115$\pm$5\,mas or 679$\pm$30\,au from the V838\,Mon position, except for directions near to the position of the companion where CO extends farther, up to a radius of 170\,mas. The maximum CO emission is not at the position of V838 Mon but about 20\,mas south-east from it. At maps obtained with uniform weighting, the emission appears clumpy, but none of the brightest CO clumps coincides with the positions of the stars (Fig.\,\ref{fig-line-maps-uni}). 

Emission of the SiO 5--4 line is nearly equally bright and extended as that of CO 2--1. On total intensity maps, however, the entire north-western part of the ejecta beyond a radius of $\approx$80 mas is missing when compared to CO. This is especially apparent in the channel maps (see Appendix\,\ref{appendix-SiO}). The region with missing SiO partially overlaps with the extended dust clump. The brightest SiO emission is, as in the case of CO, arising away from V838\,Mon, but bright CO and SiO clumps do not overlap.

Despite similar values of $E_u$ for the observed SO and SiO lines, the distribution of emission is significantly different for the two species. The SO emission is concentrated north-east from the position of V838\,Mon and only weak emission is seen west of component M. A similar emission distribution is seen in the corresponding line of $^{34}$SO, but at a much lower S/N. Part of the differences between SO and SiO distributions may be caused by a higher optical depth of the SiO line. We believe, however, that most of the differences arise due to intrinsic changes in the chemical composition within the molecular remnant. 

The blend of AlOH 7--6 and $^{13}$CO 2--1 is dominated by emission of the AlOH line. Although its map has a modest S/N, it shows a clumpy emission region whose overall size is smaller than the $^{12}$CO region. Also, unlike CO, the AlOH emission peaks north-west from V838\,Mon, between dust component B and the dust clump. 
The emission region corresponding to the blend of SO$_2$ and H$_2$S has very similar characteristics to that of AlOH. The excitation of the AlOH transition (with $E_u$ of 42.3\,K) requires much lower temperatures than the lines of SO$_2$ and H$_2$S (248.4 and 84\,K, respectively) strongly suggesting that the observed similarities in the location for the three species reflect gas chemical composition, not excitation. 

In the lower right panel of Fig.\,\ref{fig-line-maps} we present also a map produced from panchromatic visibilities from which the continuum emission was subtracted as a polynomial fit. It represents combined emission of all spectral lines present in the band, although is clearly dominated by the CO and SiO emission. Owing to a higher sensitivity, the map shows more clearly that the molecular envelope has the largest extent in directions close to the lines connecting the three continuum sources. The 3$\sigma$ contour reaches out to a radius of 184\,mas or 1086\,au. It is also apparent that the outer outline of the molecular region is not smooth, indicating a substructure possibly related to a form of hydrodynamical instability, possibly of the  Kelvin-Helmholtz type.

\begin{figure*}
    \centering
    \includegraphics[width=0.33\textwidth, trim=0 0 0 0]{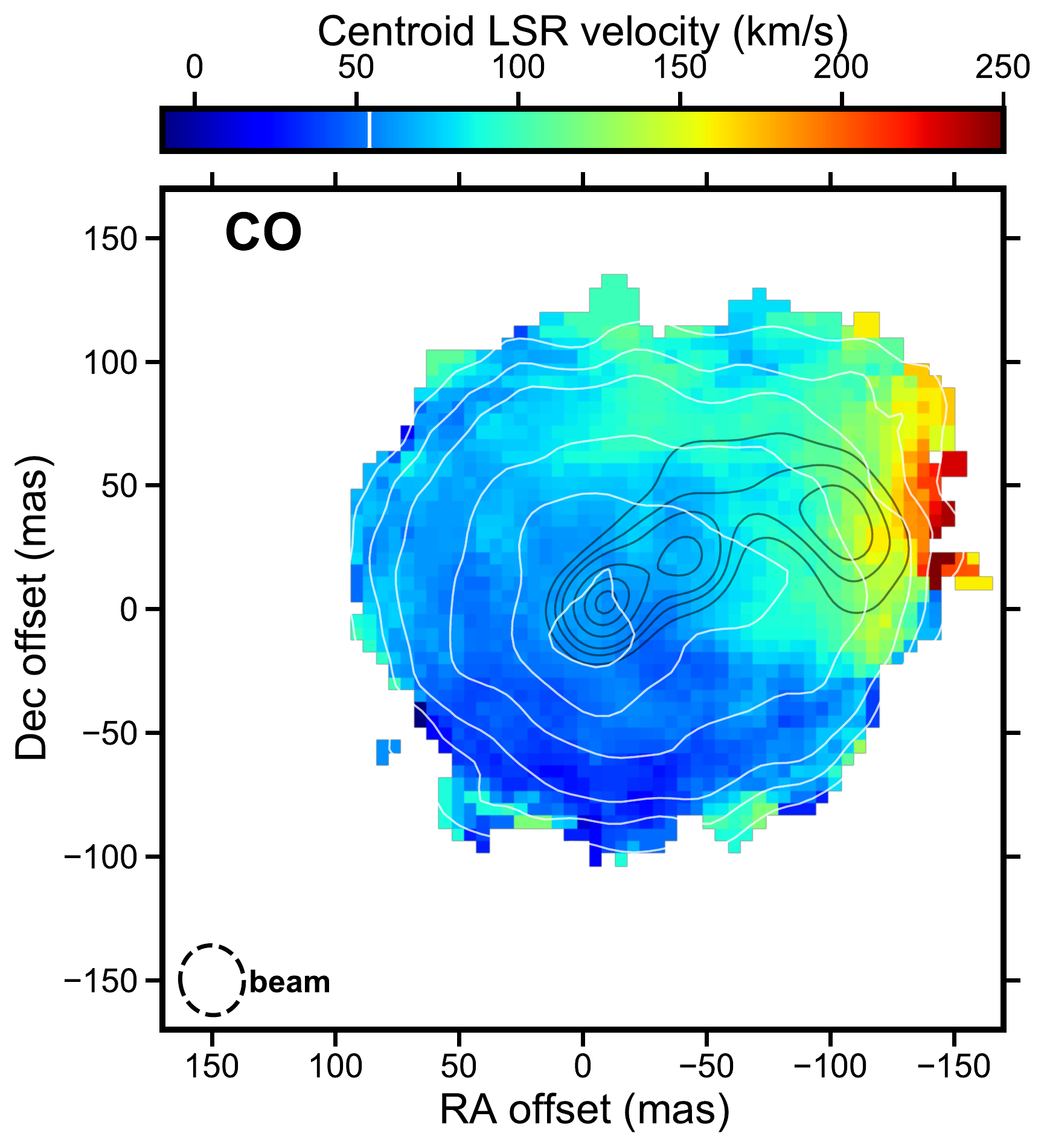}
    \includegraphics[width=0.33\textwidth, trim=0 0 0 0]{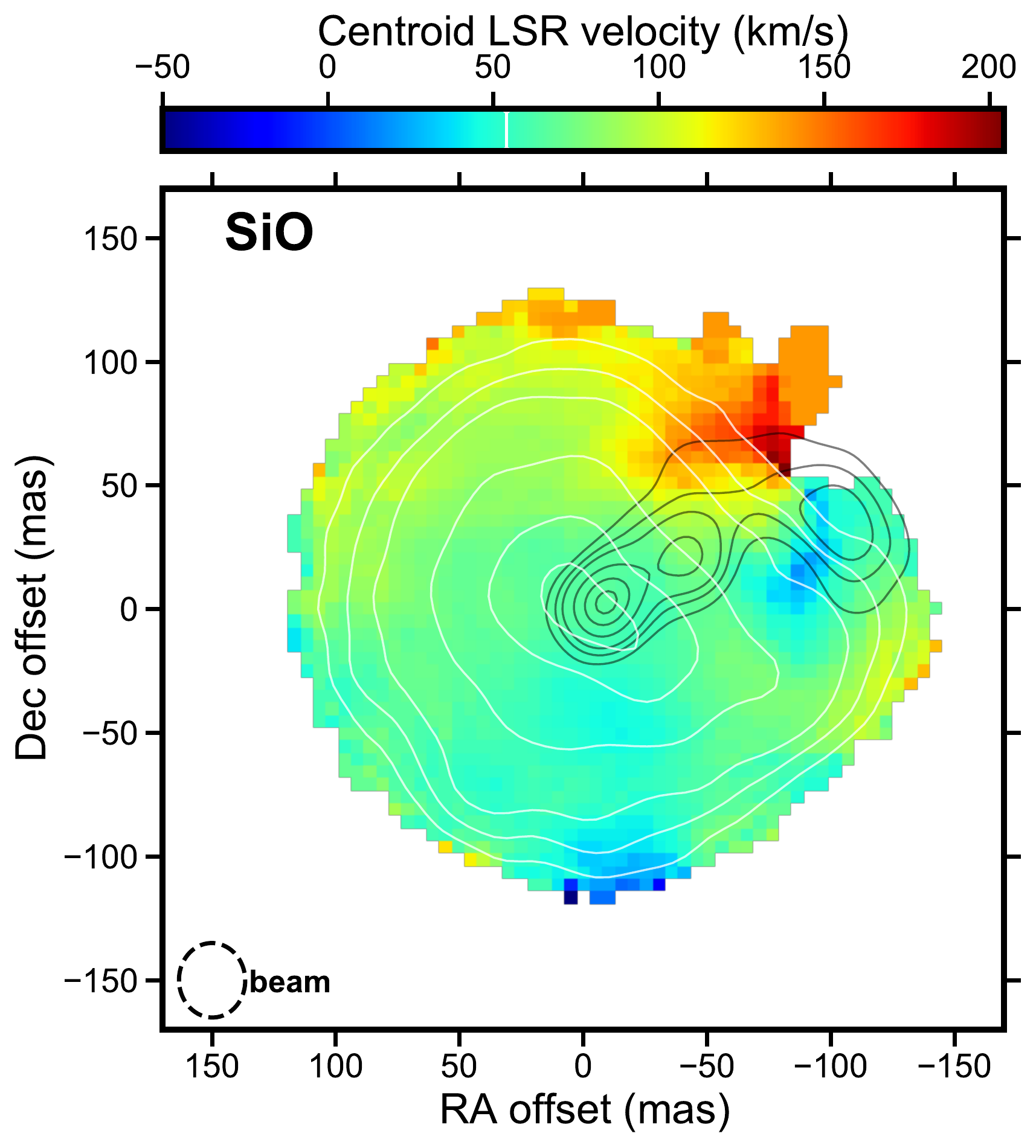}
    \includegraphics[width=0.33\textwidth, trim=0 0 0 0]{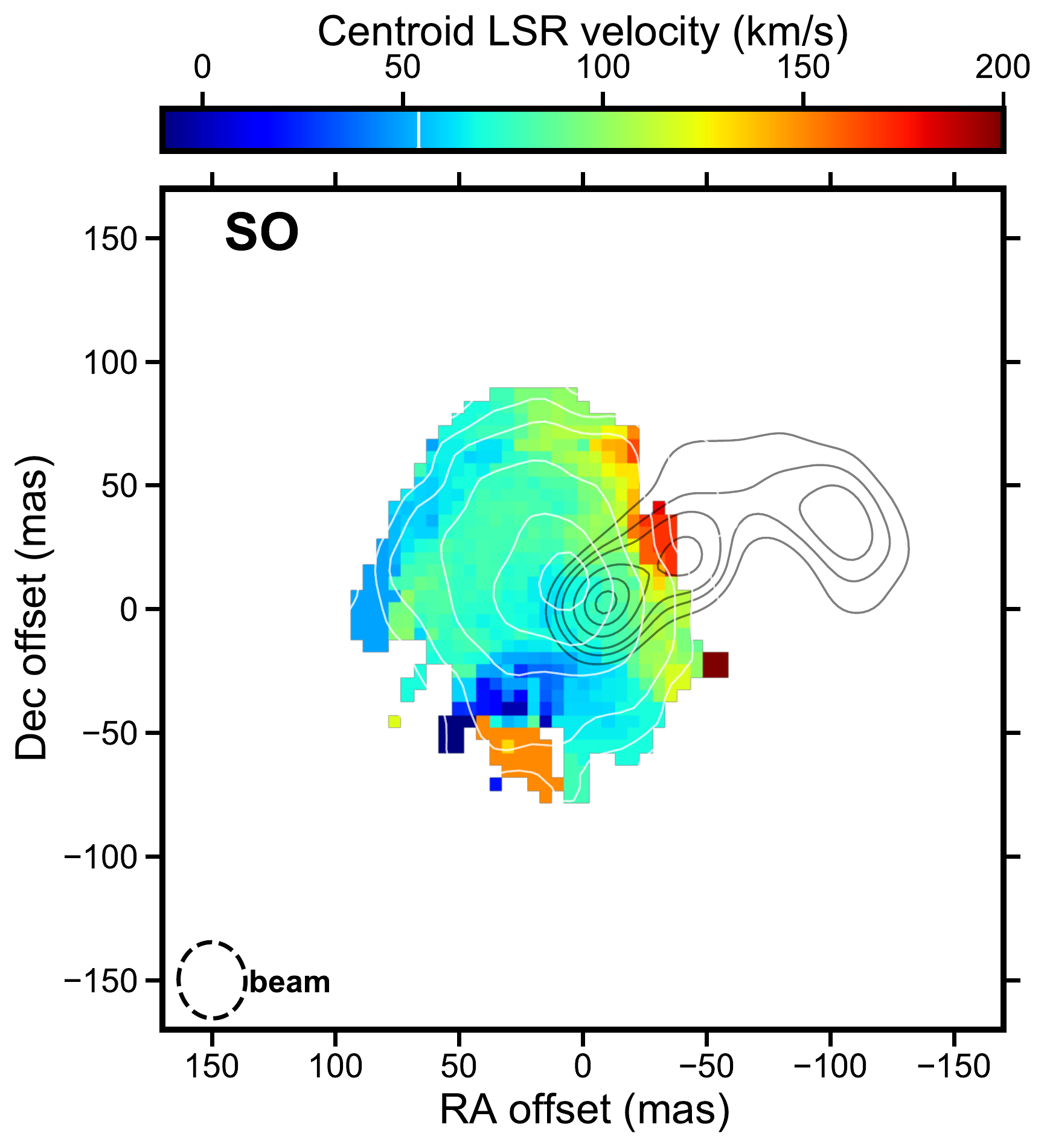}
        \caption{First-moment maps of unblended molecular lines. White contours show total intensity of the given line. Grey contours show continuum emission. Contours are drawn at 0.2, 0.3, 0.4, 0.6, 0.8, and 0.95 times the peak flux. White line at the colorbar marks the stellar (systemic) velocity of V838\,Mon. The maps were produced at natural weighting of visibilities. The moment maps were calculated after masking all pixels with total intensity below the 5$\sigma$ map noise level.}
    \label{fig-line-maps-mom1}
\end{figure*}

The velocity structure of the molecular cloud is complex. We illustrate it in Fig.\,\ref{fig-line-maps-mom1} with first-moment maps. Overall, the south-eastern part of the envelope is more blueshifted than the north-western part. The maps certainly do not show the signature expected for an expanding non-rotating spherical shell.  With respect to the systemic LSR velocity of V838\,Mon \citep[54\,\kms;][]{KamiSubmm}, significantly higher speeds are observed in the escaping (redshifted) part than in the approaching one. Highly redshifted gas is seen mostly in the north-western part of the envelope and close to its outer edges. Emission of SO probes only a small part of the envelope, but also shows an excess of redshifted gas in the western part.

\begin{figure*}
    \centering
    \includegraphics[width=0.33\textwidth, trim=0 0 0 0]{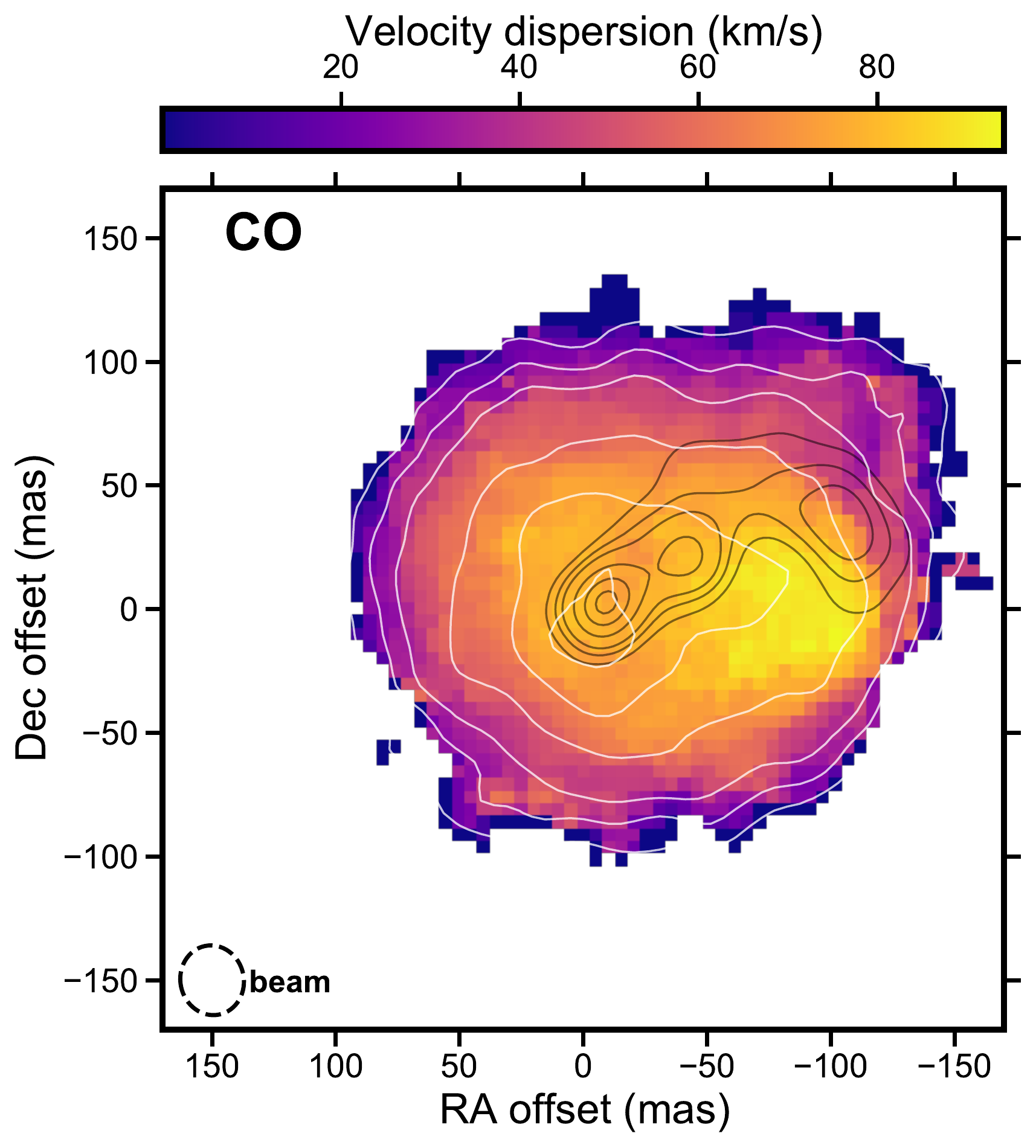}
    \includegraphics[width=0.33\textwidth, trim=0 0 0 0]{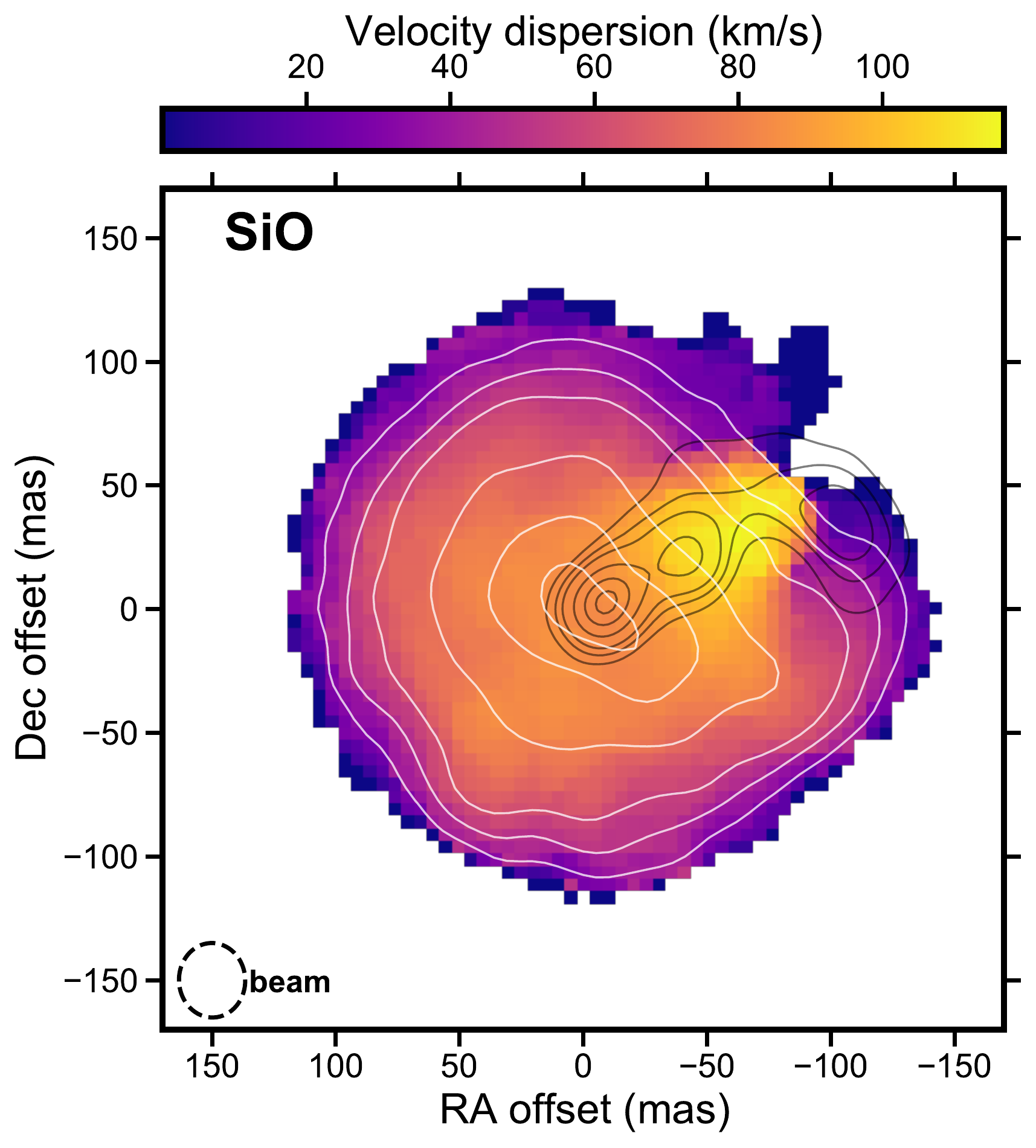}
    \includegraphics[width=0.33\textwidth, trim=0 0 0 0]{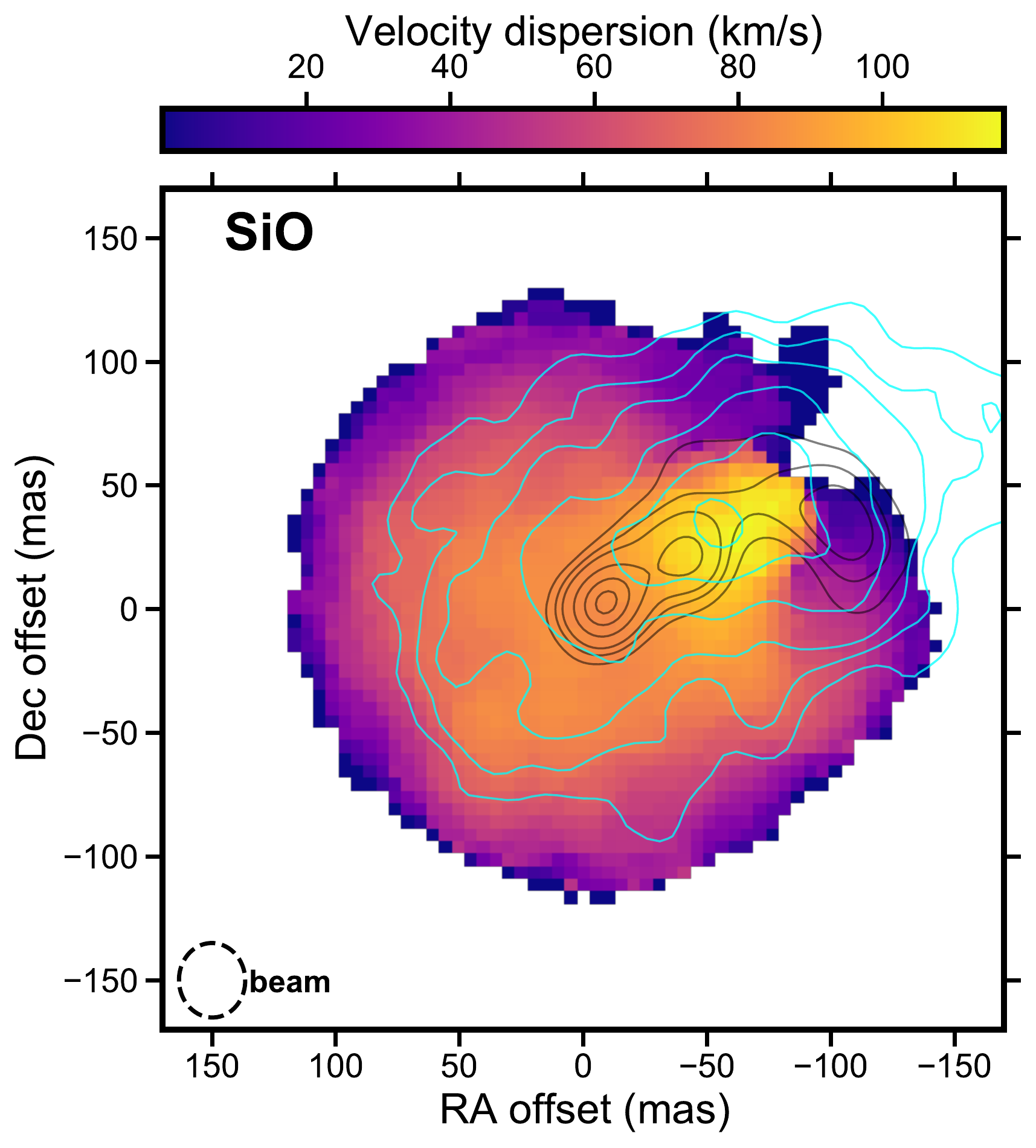}
        \caption{Second-moment maps for molecular lines of CO 2--1 and SiO 5--4. White contours show total intensity of the given line. Grey contours show continuum emission. Cyan contours in the right panel show total intensity of the SO$_2$ and H$_2$S blend. Contours are drawn at 0.2, 0.3, 0.4, 0.6, 0.8, and 0.95 times the peak flux. The maps were produced at natural weighting of visibilities. }
    \label{fig-line-maps-mom2}
\end{figure*}

The envelope kinematics can also be characterized through second moment maps, which show dispersion in radial velocity. The maps shown in Fig.\,\ref{fig-line-maps-mom2} combine the dispersion caused by the Doppler effect arising in the projected envelope and the dispersion intrinsic to the gas velocity field. For CO and SiO, circular symmetry is broken again in the western or north-western part of the second-moment maps. Assuming that most of the observed dispersion arises from the Doppler effect and that the envelope has a roughly symmetric structure, the excess dispersion of about 30\,\kms\ in the western and north-western envelope can be interpreted as a measure of excessive gas motions, possibly caused by stronger turbulence. The region with the largest dispersion seen in SiO coincides with the location of the strongest emission of the SO$_2$+H$_2$S blend (see the right panel of Fig.\,\ref{fig-line-maps-mom2}). 

\section{Optical spectroscopy near the ALMA epoch}\label{sec-optical}
Here we briefly present optical spectra of V838\,Mon from 2012 and 2020 that bracket the epoch of our ALMA observations and that are complementary to the ALMA data by presenting information about the supergiant star and its circumstellar medium on the line of sight to the bright star. While a full description and presentation of the spectra will be given in a dedicated paper, we provide the basic information on the optical spectra in Appendix\,\ref{appendix-optical}. the spectral evolution of the object at optical wavelengths prior to 2012 is discussed in depth in \citet{Loebman}, \citet{KamiKeck}, \citet{TylendaSpec2009} and references therein. 

Our flux-calibrated broadband X-SHOOTER spectrum from 2012 was used for spectral typing. The spectrum was first compared to catalog spectra of red giants and red supergiants with known spectral types and in spectral regions which are least affected by the thick circumstellar envelope of V838\,Mon. A remarkably good match was found with SV* HV 1963, a Li-rich AGB star in the Small Magellanic Cloud. Its spectrum was found in the X-SHOOTER Spectral Library \citep{xshooterStellarLibrary}. It is classified as M4 Ib--II \citep[after][]{class1}, but detailed models of the star yield $T_{\rm eff}$=3300--3350\,K and $\log g$=--0.027 to 0.0 and [Fe/H]=--0.5 \citep{Plez,Kiselman}. We next compared the spectrum of V838\,Mon to synthetic MARCS spectra \citep{marcs} and found a good correspondence between 3300\,K stellar models and molecular bands of V838\,Mon originating from the excited vibrational or electronic states. The same comparison to synthetic spectra was performed for the 2020 spectrum from SALT. It yielded a slightly higher photospheric temperature, of 3500\,K, and could suggest a slow rise in the photospheric temperature of V838\,Mon over the last decade or so. These temperatures are significantly higher than 2000--2200\,K claimed in \citet{Loebman} and based on optical and infrared (IR) spectra obtained a few years earlier. Possibly, the earlier study did not exclude spectral regions dominated by molecular absorption of cool non-photospheric gas and their estimate does not correspond to the actual photosphere of the star. The temperatures of 3300 and 3500\,K are more consistent with those of \citet{KamiKeck} (3200\,K) and \citet{TylendaEngulf} (3270\,K) in earlier epochs (2005 and 2009, respectively).

\begin{figure}
    \centering
    \includegraphics[width=0.99\columnwidth, trim=0 0 0 0]{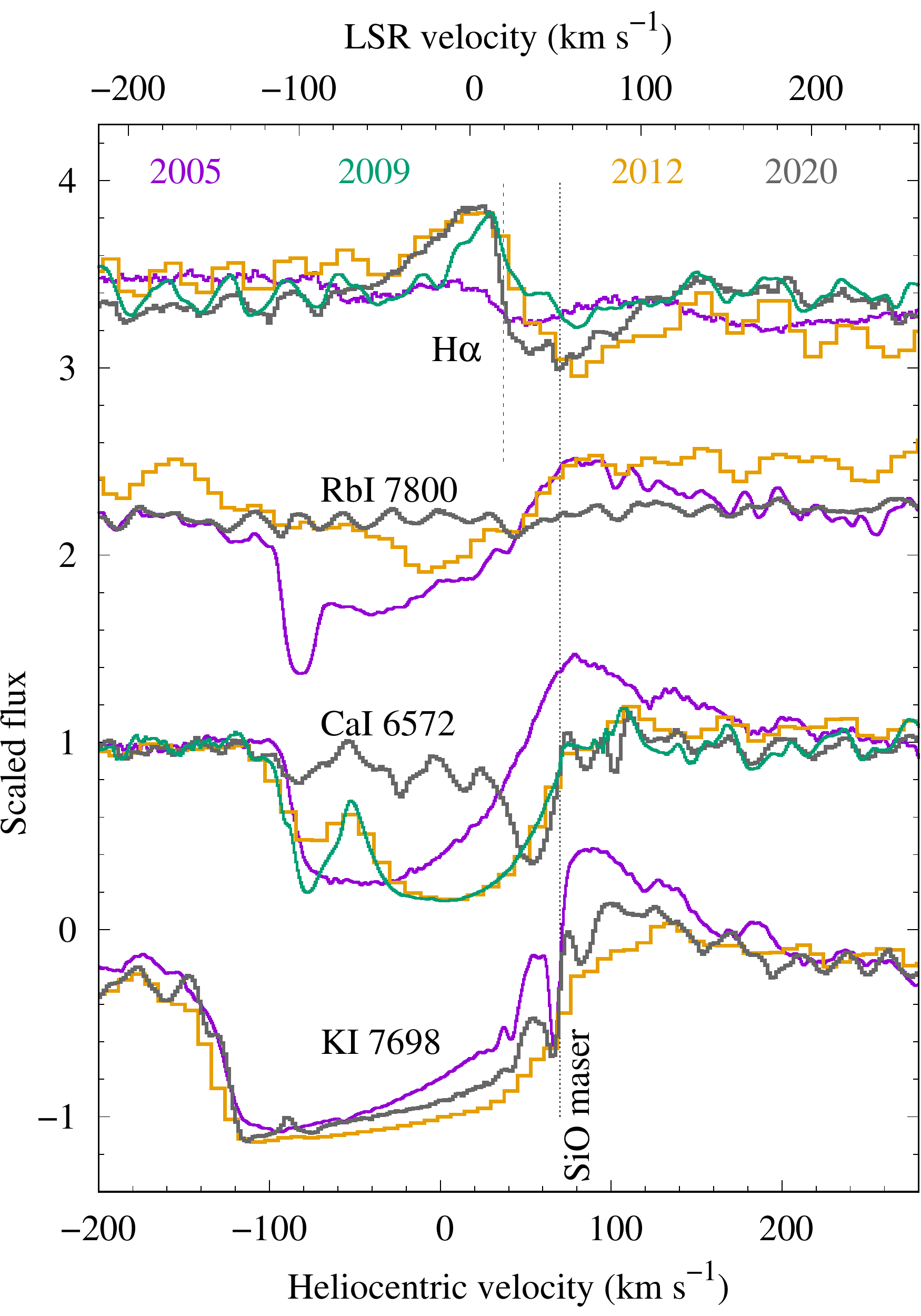}
        \caption{Temporal changes in selected atomic line profiles of V838\,Mon. Keck/HIRES spectra from 2005 \citep{KamiKeck} are shown in purple, VLT/UVES spectra from 2009 in green \citep{TylendaSpec2009}, VLT/X-SHOOTER spectra from 2012 in orange, and SALT/HRS spectra from 2020 in gray. The spectra were arbitrarily scaled to show changes in the four labeled atomic lines. The vertical lines mark the velocity of the SiO maser (or the supposed stellar systemic velocity; 53\,\kms\ LSR or 71\,\kms\ in the heliocentric frame) and the position of the turnover in the inverse P-Cyg profile of H$\alpha$ (37\,\kms\ heliocentric).}
    \label{fig-optical}
\end{figure}

Our spectra show strong signatures of the cool circumstellar matter on the line of sight, mainly in lines of neutral alkali metals and in molecular bands. Sample atomic profiles for four epochs are shown in Fig.\,\ref{fig-optical}. The strongest atomic lines are resonance transitions of abundant species, such as the optical doublets of \ion{K}{I} and \ion{Na}{I}, or the $\lambda$5110 line of \ion{Fe}{I}. These are the broadest absorption features spreading from the Local Standard of Rest (LSR) systemic velocity of the star of 53\,\kms\  \citep[cf.][]{KamiSubmm} down to LSR velocities of $\approx$--150\,\kms. The deep absorption lines did not change significantly over the last years. 
In weaker atomic lines, the same velocity range is occupied by several narrower features which become weaker with time, as illustrated in Fig.\,\ref{fig-optical} in profiles of \ion{Rb}{I} $\lambda$7800 and \ion{Ca}{I} $\lambda$6572.
The absorption has completely disappeared in some species in the most recent spectrum, for instance in the \ion{Rb}{I} line. The continuing fainting of the weaker absorption lines, known from earlier spectroscopic monitoring of V838\,Mon, shows primarily the expanding merger ejecta that becomes more diluted and cooler with time. Absorption components at lower radial velocities, close to the systemic one, most likely represent an ongoing wind originating in the supergiant-like stellar remnant. In the 2020 SALT spectrum lines from highly excited states, such as lines of \ion{Ti}{I} from a 0.8\,eV state, display a relatively narrow absorption component at a heliocentric velocity of +85\kms\ indicative of an infall. Its location and origin is unclear, but it is very likely it coexists with the wind component in the immediate vicinity of the photosphere. 

Among all atomic features, the profile of H$\alpha$ is unique (Fig.\,\ref{fig-optical}, top). It shows a weak inverse P-Cyg type profile with the absorption component centered nearly at the systemic velocity. The H$\alpha$ emission peaks near an LSR velocity of about 5\,\kms. The origin of the H$\alpha$ feature has been unclear, but it was speculated it may arise in the medium photoionized by the B star \citep{Loebman,TylendaSpec2009}. Our recent spectra show it is a persistent feature on time scales longer than a decade and is relatively constant both in position and in intensity.

Overall, referring to gas ejected in the merger event, the optical spectra from before and after the epoch of ALMA observations show the remnant has been in a similar state since around 2005, that is, since the disappearance of the B star in the visual and since pure emission lines vanished.

\section{3D reconstruction of the remnant}\label{sec-3d}
\subsection{Dust}\label{sec-3d-dust}
We attempted a reconstruction of dust distribution in the remnant of V838\,Mon using RADMC-3D\footnote{\url{ https://www.ita.uni-heidelberg.de/~dullemond/software/radmc-3d/index.php}} \citep{radmc3d}. For a given configuration of a dust density distribution and location of radiation sources, dust temperature is calculated using the radiative equilibrium assumption. Next, a ray-tracing method is used to simulate images and integrated fluxes of the system at different wavelengths. We compared the simulations to a spectral energy distribution of V838\,Mon ranging from the $U$ optical band to the millimeter band of ALMA. The data sources for the observed SED are briefly described in Appendix\,\ref{appendinx-a}. Simultaneously, we tried to reproduce those spatial features seen in the ALMA maps that can be clearly distinguished from each other. 

\begin{figure}
    \centering
    \sidecaption
    \includegraphics[width=0.95\columnwidth]{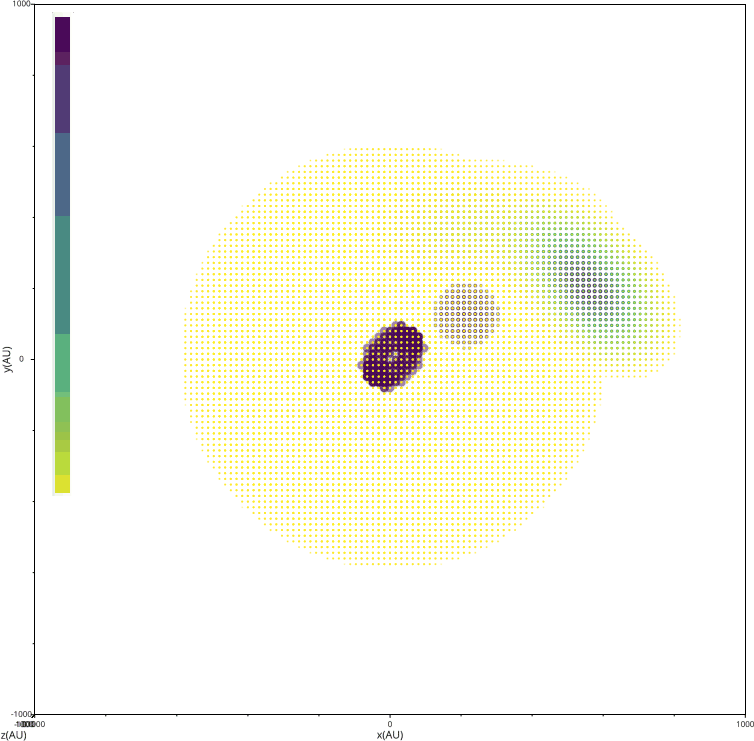}
    \caption{3D model of dust distribution in V838\,Mon in 2019. An animation presenting the model is available  at \protect\url{https://drive.google.com/file/d/1miIauR3Ie7PXiR4by49fz8w2WayJpDqV/view?usp=sharing}. Model grid points are represented by semi-transparent spheres whose size and color correspond to local dust density. The color scale is linear from $1.0 \cdot 10^{-19}$ g\,cm$^{-3}$ (yellow) to $1.6 \cdot 10^{-16}$ g\,cm$^{-3}$ (dark purple). The sky plane is for $x=0$ and $y=0$ and the $z$ axis is along the line of sight. The volume shown is a cube of a size of 2000\,au.}\label{fig-model3D}
\end{figure}

\begin{figure}
    \centering
    \sidecaption
    \includegraphics[width=0.99\columnwidth, trim=0 0 0 0]{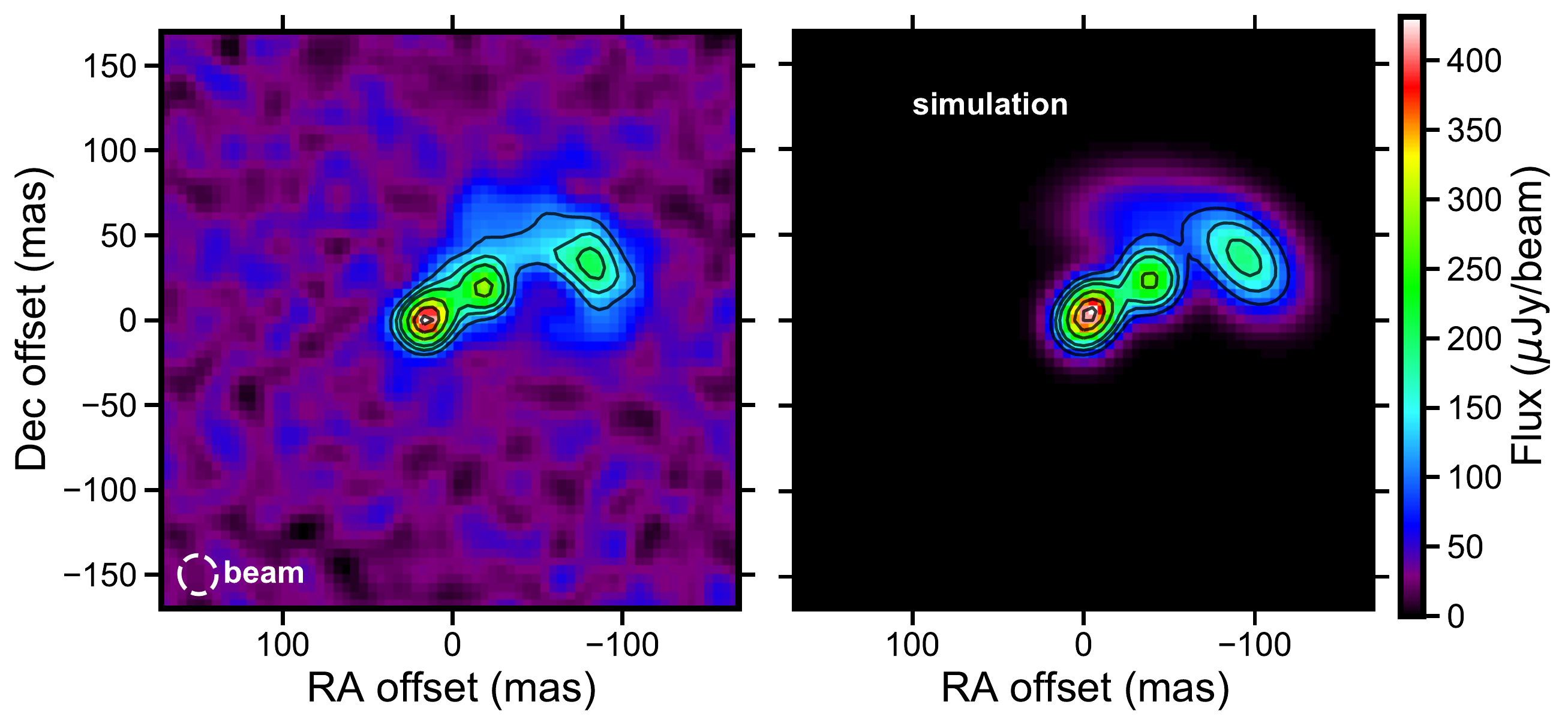}
    \caption{Comparison of observed (left) and simulated (right) maps of continuum emission in ALMA band 6. Both images are displayed with the same color scale, contour levels, and angular resolution.}\label{fig-compare}
\end{figure}

\begin{figure}
    \centering
    \sidecaption
    \includegraphics[width=0.79\columnwidth, trim=0 0 0 0]{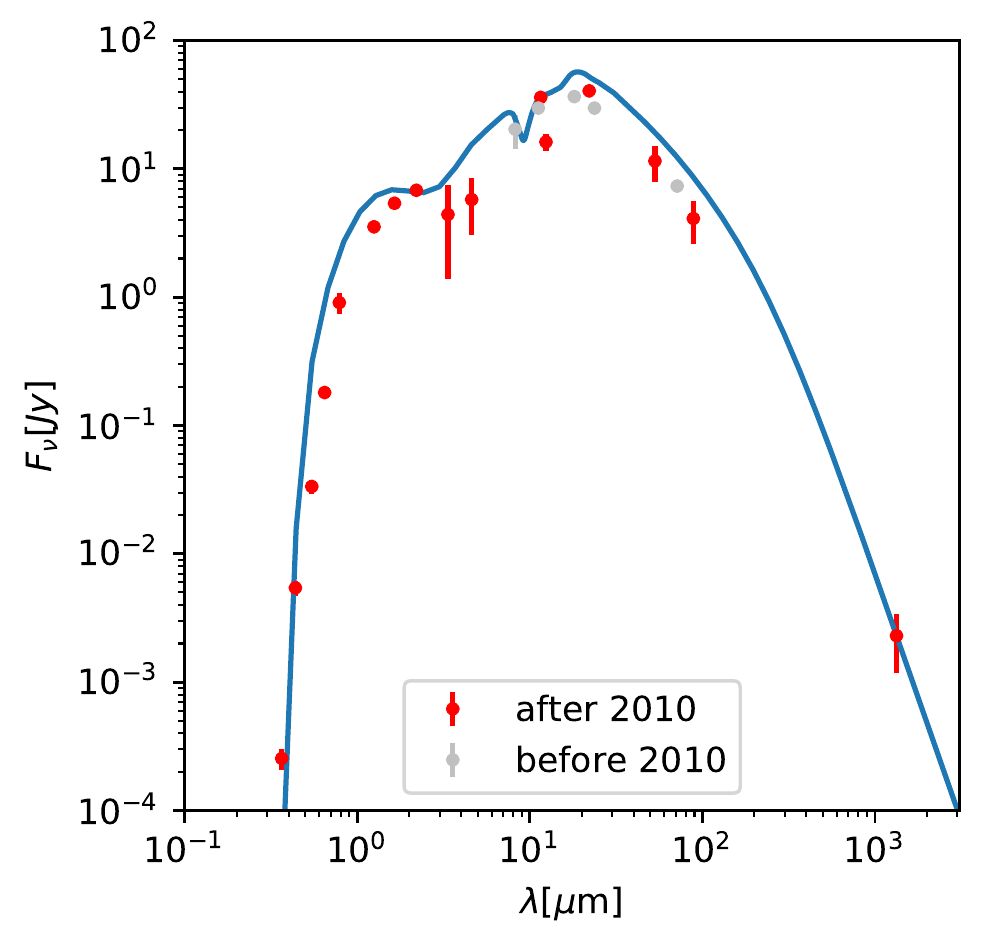}
    \caption{Comparison of the observed (points) and simulated (line) SED. Uncertainties in dust opacities are much larger than the apparent discrepancies between the observations and the simulation.}\label{fig-sed}
\end{figure}

Even in the simplest considered configuration, the modeled system has numerous free parameters. They cannot be uniquely constrained by our observations and therefore the model we attempted to match to the observations was rather exploratory. After running thousands of simulations, we arrived at a model described below and shown in Fig.\,\ref{fig-model3D}. A comparison between observed and simulated images and SEDs are shown in Figs.\,\ref{fig-compare} and \ref{fig-sed}.
\begin{itemize}
    \item The only heat sources in the system are the two stars. The B3\,V companion was represented by a black body of a temperature of 18\,kK, a radius of 3.1$\cdot$10$^{11}$\,cm (4.5\,R$_{\sun}$), and a luminosity of 1.9$\cdot$10$^3$\,L$_{\sun}$. These values correspond to a spectral type of B3\,V. We located the blue star in the sky plane, 230\,au away from V838\,Mon, that is at the same distance along the line of sight as V838\,Mon. 
    V838\,Mon was represented by a black body of a fixed temperature of 3300\,K, consistent with spectral typing performed on the most recent optical and IR spectra of the star (Sect.\,\ref{sec-optical}). The radius and thus luminosity of V838\,Mon were adjusted to match the SED. The optimal values we arrived at are 3.25$\cdot$10$^{13}$\,cm (464\,R$_{\sun}$) for the radius and 2.3$\cdot$10$^4$\,L$_{\sun}$ for the luminosity. 
    \item The biggest uncertainty in the SED reconstruction comes from dust opacities as a function of wavelength and location in the remnant. Since the gas seen in absorption and emission spectra is oxygen-rich (that is, it is indicative of a higher abundance of oxygen than of carbon), a form of silicate dust is expected to dominate in the opacity curve. We arbitrarily implemented an opacity curve of Mg-rich silicates (amorphous pyroxene with 70\% Mg-rich and 30\% Fe-rich) compiled from data in \citet{silicate1} and \citet{silicate2} and for a single grain size of 0.1\,$\mu$m. Grain density was set to 3\,g\,cm$^{-3}$. This opacity curve matches slightly better the IR part of the SED near 10\,$\mu$m than most other silicate opacities popularly used in the modeling of circumstellar media. Material in the real remnant probably comprises a range of dust species with different grain sizes and compositions, but those cannot be currently uniquely constrained.
    \item The simulated M component is a disk-like or torus-like structure immediately surrounding the stellar remnant of V838\,Mon and has a dust mass of 9.4$\cdot$10$^{-4}$\,M$_{\sun}$. It was implemented as a structure filled with dust for a range of azimuthal angles, 65\degr<$\theta$<115\degr, and whose plane is inclined to the line of sight by 26\degr\ so that the disk edge does not obscure the central star for our line of sight. The structure has a maximal radius of 95\,au. It is hollow in the center with a hole radius of 60\,au. At the outer hole radius, the dust temperature is 1350\,K. The hole can be interpreted as a clearing related to the sublimation of dust. When inclined, such an elongated structure explains well the shape of the beam-deconvolved component M (Sect.\,\ref{sec-obs}; Fig.\,\ref{fig-line-maps}). It is also consistent with VLTI observations of the region at MIR wavelengths \citep[][Mobeen et al., in prep.]{Olivier}. The dust density was assumed to increase with distance from V838\,Mon as $r^{0.75}$. We also attempted to reproduce the M component by a spherical dusty cloud, such as one expected from a spherically expanding supergiant wind. However, such a configuration results in a higher extinction towards the central star than observed. We favor models with a flattened structure, such as a disk or a torus.
    \item External to the disk we place a spherical shell at radii between 95 and 600\,au, which we refer to as a mergeburst shell. Although extended, this structure was implemented to have a low dust density, on the order of 10$^{-19}$\,g\,cm$^{-3}$, and which only slightly increases with distance (as $r^{0.1}$). The dust mass of the structure is rather low, 8.0$\cdot$10$^{-5}$\,M$_{\sun}$. We associate the shell with the merger ejecta directly seen in the molecular emission traced by ALMA and seen in absorption lines of neutral metals and  of simple molecules at optical wavelengths throughout the entire post-outburst evolution of the object (cf. Sect.\,\ref{sec-optical}). In the model, the spherical mergeburst shell is the only source of extinction on the line of sight of the cool star, with $A_V\approx$1\,mag.
    \item We surround the blue star with a shell of dust of a constant density (for simplicity) of 1.5$\cdot$10$^{-16}$\,g\,cm$^{-3}$ and extending from 10 to 95\,au from the hot star. Again, the inner clearing is caused by dust sublimation. The outer radius is constrained by the beam-deconvolved size of component B. We require this extra shell in our model to produce the submm source, but it is not clear if the dusty structure is limited to a spherical region directly surrounding the B star or is part of a more extended component whose parts, not illuminated by the stars, are not visible in the map. In principle, we could increase the local density of the mergeburst shell, so that it produces the observed submm signature around star B, but then the extinction toward the red star becomes much higher, so it becomes inconsistent with visual observations. Alternatively, a flattened structure surrounding V838\,Mon and overlapping the orbit of the B star may be responsible for component B. The dust mass of the small shell surrounding solely the B star is 9.4$\cdot$10$^{-4}$\,M$_{\sun}$ but if component B is associated with an extended but dark dust, the actual mass may be much higher.
    \item As the last component of the simulation, we added two Gaussian-shaped clouds to reproduce the extended submm continuum emission north and north-west from the B component. Gaussian shapes are chosen to ease the implementation, and we only aimed at reproducing the main features of the extended component. (In fact, reproducing exactly the ALMA images for this extended component would be trivial.) We locate the two representative clouds in the sky plane, that is in the same plane as the two stars. The relative location of the components is important as the shell surrounding component B casts a shadow on the two Gaussian extended clouds partially blocking light from the bright cool star. The clouds are only excited by photons of the two stars. No inner heat source, such as shocks, were implemented, although they may be adequate for this spatial component (see Sect.\,\ref{sec-discussion}). The cumulative mass of the dispersed dust in these two clouds is 6.1$\cdot$10$^{-3}$\,M$_{\sun}$. This mass is thus significantly higher than that of all the other components of the system. The high mass is required to produce the submm fluxes in the absence of an internal heat source.    
\end{itemize}

\subsection{Molecular gas}\label{sec-3d-mol}
We also attempted to reconstruct with RADMC-3D the distribution of molecular gas. Such models are however even less constrained than those for dust because gas models require extra input for the temperature and velocity structure (and thus even more free parameters). Although we were able to construct models that very satisfactorily reproduce maps and spectral profiles of CO observed by ALMA, the solutions we find are by no means unique or particularly revealing. Observations of at least one more transition of CO or SiO at a similar angular resolution would lift many of the model degeneracies. For CO, a single nearly-spherical shell with temperature, density, and radial velocity gradients given as power laws was sufficient to explain most of the observational characteristics. We also found that CO line profiles are better reproduced with the systemic velocity of +70\,\kms\, rather than 54\,\kms\ defined by the SiO maser near V838\,Mon.

\begin{figure}
    \centering
    \sidecaption
    \includegraphics[width=0.99\columnwidth, trim=0 0 0 0]{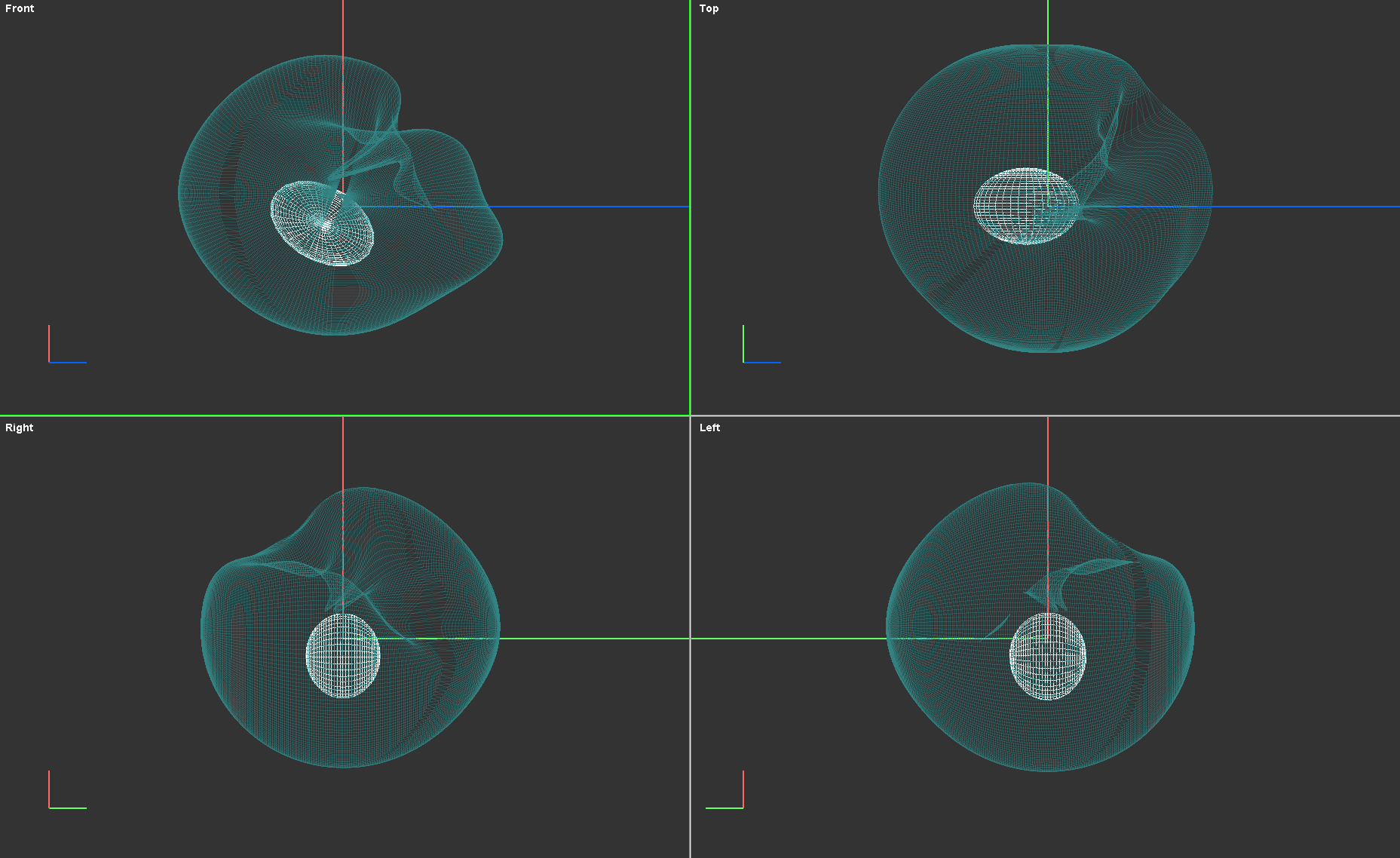}
    \caption{Model of spatial extent of SiO emission around V838\,Mon. The mesh structures show two regions with different velocity fields and density (or intensity) power laws which reproduce channel maps and position-velocity diagrams of the SiO line. The outer shell contains a characteristic void which the model attempts to reproduce. The "front" view (top left) corresponds to that of the terrestrial observer. 
    }\label{fig-SiOshape}
\end{figure}

To investigate the possible structure of the outflow in molecular tracers other than CO, we used the Shape software\footnote{\url{https://wsteffen75.wixsite.com/website}} \citep{shape1}, which gives more flexibility in constructing complex spatial models than RADMC-3D. We focused on reproducing the observations in optically thin limit with very simplified radiative transfer. Here we present only the results for the SiO transition. It required two shell-like structures, each with a different density profile (or rather intensity) and slightly different velocity fields. The latter were implemented as simple relations with radial velocity increasing linearly with distance ($\varv=kr$). The velocity gradient is steeper for the outer structure ($k$=1.9) than for the inner one ($k$=1.7). The inner shell was best reproduced as an elliptical structure whose center is slightly shifted with respect to the position of component M. The larger structure is more spherically symmetric, but it required introducing a cone-shaped void in the north-western part. The rather complex shape of the larger structure is presented in Fig.\,\ref{fig-SiOshape}. The void seems to extend from the position of the B component and grows in size with distance from the geometrical center. Our simple model satisfactorily reproduces structures in channel maps, shown in Appendix\,\ref{appendix-SiO}, and position-velocity diagrams but no attempt to reproduce absolute intensities was undertaken in this case. The model strongly suggests that the void is directly related to the passage of matter near component B, but does not shed light on whether it is caused by depletion of SiO or by a change in excitation conditions.

\section{Discussion}\label{sec-discussion}
The ALMA view on the 2019 remnant clarifies the architecture of the system 17\,yr after the merger and for the first time shows directly the location of the B3\,V companion.  The different system components identified in the observations and through the modeling pose however also several problems. These are discussed below, first for the gaseous molecular component.   
\subsection{Molecular gas in the remnant of V838\,Mon}\label{sec-discussion-gas}
\paragraph{Historical view}
So far, most of the information about the circumstellar environment of the V838\,Mon system has come from single-line optical and infrared spectroscopy. During the outburst, the maximal outflow velocities traced in atomic features were about 500\,\kms\ \citep{KipperConf,WisniewskiSpecpol} and the outflow velocities were considerably changing over the three months of the outburst \citep{Munari2002, KipperKlochkova, KipperConf, WisniewskiSpecpol, RushtonEvol, CrauseSpec}. 
Broad absorption lines from cool atomic and molecular gas have been observed since the end of the eruption, first as part of P-Cygni profiles \citep{KipperKlochkova, KipperConf, KamiKeck,TylendaEngulf, Geballe, Lynch2004, WisniewskiSpecpol} and later on, after around 2006, mainly as pure absorption features \citep{Loebman, KamiSubmm, Geballe}. The absorption troughs indicated maximal outflow velocities of 250\,\kms\ \citep{KamiKeck, TylendaEngulf} but most of the material was moving at lower speeds. In particular, \cite{TylendaEngulf} found three slower atomic components, with outflow velocities of about 50, 110, and 150\,\kms, and \citet{Geballe} found similar CO components at 15, 85, 150\,\kms. Since the gas ejection velocities were evidently changing during the eruption, the remnant was expected to be stratified. Spherically symmetric models with homogeneous and isothermal shells were adopted for the circumstellar molecular gas \citep{Lynch2004,Lynch2007,Loebman}. Presenting for the first time maps of the molecular emission, we can conclude that the ejecta is not as much stratified as it is clumpy (Fig.\,\ref{fig-line-maps-uni}). Thus, absorption spectroscopy alone, limited to one line of sight, does not provide a full picture of the entire remnant. 

\paragraph{Wind}
Spectral signatures at optical and submm wavelengths reveal an ongoing wind that is emanating from the central red supergiant. Based on the most recent optical high-resolution spectra from the SALT telescope (Sect.\,\ref{sec-optical}), the terminal velocity of the wind is $\lesssim$50\,\kms\ \citep[cf.][]{TylendaEngulf,Ortiz-Leon,KamiSubmm}. Assuming that the wind took the present form in mid 2002 and freely expanded over the 17\,yr preceding the ALMA epoch, it should have reached a radius of 179\,au or 30\,mas, which constitutes a third of the size of the remnant traced by CO emission in the ALMA epoch (see right panel of Fig.\,\ref{fig-line-maps}). If this is correct, the wind has not yet reached the orbit of the companion. The SiO maser, which appears to gradually drop in intensity \citep{Ortiz-Leon}, is excited in the innermost regions of this clumpy wind. The emission of different species traced by ALMA, especially emission of SO (Fig.\,\ref{fig-line-maps}), shows excitation variations or chemical inhomogeneities within the wind, not unlike those known from genuine red supergiants \citep{KamiVYSurvey,adande,muCeph}. Since the stellar remnant is expected to be a highly magnetic object \citep{Sokermagnetic,schneider}, perhaps some of these asymmetries can be linked to magnetism.

\paragraph{Merger ejecta}
ALMA provides the clearest view so far on the molecular merger ejecta. The molecular gas seen in CO ro-vibrational absorption bands had the highest velocities of $\approx$190\,\kms\ \citep{Geballe}, much lower than those observed in atomic gas. The warm molecular component at this speed should have reached a radius of 700\,au or 119\,mas in the ALMA epoch. This size is in excellent agreement with the observed tangential extent of the rotational CO emission, that is, to nominal radii of 115\,mas. This suggests that the tangential motions of the bulk of molecular gas are of the same magnitude as those along the line of sight. The merger ejecta is roughly spherical. 
The full width of the CO $J$=2--1 line profile representing the entire molecular cloud mapped by ALMA indicates maximal radial velocities of 230$\pm$20\,\kms, which are higher than those seen in the IR. This high-velocity gas is seen as weak emission in the line wings and has very low surface brightness so that it could have been missed in the IR observations. 

\paragraph{Unseen atomic gas}
Assuming spherical symmetry and ignoring the wind component, the merger ejecta should have taken the form of an expanding shell whose outer parts contain the fastest ejecta and the inner ones have the lowest outflow velocities associated with the eruption. 
The fastest ejecta (at 500\,\kms) should have reached a distance of 1840\,au or 311\,mas in mid 2019, much farther away than any molecular or dust tracer seen by ALMA. Our observations therefore show us only the inner ($\sim$half) part of the gaseous remnant. Observations with an instrument capable of resolving atomic or dust-scattered emission at scales of 600\,mas would be necessary to probe the high-velocity ejecta. This should be possible, for instance, with the SPHERE instrument on the Very Large Telescope. Unfortunately, emission lines are currently very weak in V838\,Mon. 
(Sect.\,\ref{sec-optical}; \citealp[see also][]{Loebman}). 

\paragraph{No molecular disk?}
With the limited molecular data in hand, we find no direct evidence for a disk or a flattened structure that is known from earlier IR observations and is hinted by our dust continuum observations (Sect.\,\ref{sec-3d-dust}). Perhaps no suitable tracer has been covered by our ALMA spectra.

\paragraph{Interaction with the B star}
Gas kinematics revealed by ALMA show the north-western region in the vicinity of the B component to be the most turbulent and the most redshifted portion of the remnant, suggestive of gas interactions with the B star. The possible nature of the interaction is discussed in Sect.\,\ref{sec-discussion-interaction}. The different molecular tracers observed by ALMA show also chemical composition variations within the inner remnant (SO vs. CO) and in the outer merger ejecta. These variations are most pronounced near (or, from V838 Mon's perspective, "behind") the B companion. Although there is a remote chance that the observed differences in emission distribution are related to different excitation conditions, we find it more likely that this turbulent region hosts shocks that strongly influence gas molecular composition. Some species appear to have enhanced abundances in the interaction region, including SO$_2$/H$_2$S and AlOH, others are depleted (SiO, Sect.\,\ref{sec-3d-mol}). In astrochemical studies of the interstellar medium (ISM), S-bearing species are often assumed to trace shocks \citep{sulfurshocks,sulfurNew}, corroborating our interpretation. The interaction of the merger ejecta with the B-type star and its environment poses an interesting case for circumstellar shock-induced chemistry worth further investigations. If shock chemistry were influencing the chemistry in this interaction region, then, on face value, the lack of SiO would be surprising. 
Since the seminal work of \citet{Schilke1997}, 
an enhanced SiO abundance is recognized as a tracer of shocks in the ISM. These authors ascribe an enhancement of the SiO abundance, which is generally low in the ISM, to the sputtering of silicate dust grains, a process in which atomic silicon is set free and oxidized.


Let us assume that the asymmetry in the first-moment maps of molecular emission is caused by an interaction of the outflow with the companion. Then the redshifted gas can be interpreted as part of the material that was slingshot away from us by the gravity of the B star. Redshifted material thus indicates that the companion is not exactly in the plane of the sky, but is more distant than V838\,Mon.

\subsection{Dusty remnant}\label{sec-discussion-dust}
The continuum emission components surrounding the stars can be easily explained as thermal dust emission of grains heated by the radiation of the two stars. Although it would be tempting to assume that the dust has the same spatial distribution as the molecular gas, this is evidently not the case in V838\,Mon. 

The merger remnant is surrounded by a wind which may have a dust component. The emission of this component must be rather optically thin and cannot be easily identified in the observations. Our modeling shows that the elongated component M surrounding the stellar remnant is most likely a torus-like structure seen at a moderate inclination. Although this interpretation is not firm, as our data lack the angular resolution to be definitive, such a structure is expected to exist around merged systems. Also, disks are observed in other red-nova remnants \citep{spectropolarimetry,Tylenda2011}. Such a disk or a torus-like structure is expected to carry most of the angular momentum of the former binary and consist of the common envelope material. Future interferometric observations in the IR may measure that angular momentum and thereby constrain the common envelope physics.

The nature of component B is at a first sight easy to understand. ALMA maps identify the position of the B star and leave no doubt that the star forms a physical system with V838\,Mon. Assuming that the B star is at a circular orbit of a size of 230\,au in a system of a mass of 8.0+8.4\,M$_{\sun}$, its orbital period is 1000\,yr and the orbital velocity is 3.8\,\kms. Such a wide binary of two intermediate-mass stars is expected to be eccentric \citep{Tokovinin}, but there is currently no way to constrain this orbital parameter. As mentioned before, from the timing of the interaction of the merger ejecta with the B star \citep{TylendaEngulf}, it seems that the plane of the orbit must be very close to the sky plane, but the asymmetry in profiles of molecular lines suggests that the B star is slightly more distant than V838\,Mon. It would be interesting to know if the real orbit is coplanar with the disk found in component M and, if not, as it seems to be the case, what a role did Kozai-Lidov cycles \citep{Kozai,Lidov} played in the merger of the inner binary \citep[cf.][]{Naoz}. 

It is natural to assume that the dust giving rise to the submm emission of component B comes from the dusty merger ejecta encircling the B star. Our radiative transfer modeling suggests however that, at least for a spherically symmetric merger ejecta, component B requires an enhanced dust density to produce the observed mm signature (Sect.\,\ref{sec-3d-dust}). Alternatively, it is possible that there is a ring of dust overlapping with the orbit of the B star, not crossing the line of sight, and whose major part is dark for us. One other possibility is that some micro-physical processes increase dust condensation only near the hot star, for instance, due to a collision of the ejecta with the wind of B star  \citep[cf.][]{dustWR}. This would partially explain the depletion of SiO (Sect.\,\ref{sec-3d-mol}), which in this scenario would be easily incorporated into solid material. Yet another interpretation is that some material was captured from the flow and remains bound to the star. This scenario can be, however, excluded based on the hydrodynamical considerations outlined below. Future observations of the eventual re-appearance of the B star in the optical may shed more light on the origin of component B.  

The third extended continuum source is most difficult to explain, especially because it does not have a counterpart in CO emission. It either requires a strongly enhanced dust density for a range of azimuthal angles in the outer remnant or involves an inner dispersed heat source, such as shocks and their radiation. In either case, some special conditions must have prevailed in the north-western part of the remnant. The location of the extended component is close to the line connecting V838\,Mon with component B, leading us to believe that it is the interaction of the ejecta with the B star that created the extended cloud. 

\subsection{Interaction of the merger ejecta with the B star}\label{sec-discussion-interaction}
\subsubsection{Wind accretion (deflection)}
The ejecta passing the B companion interacts gravitationally with the star, resulting in a geometrical configuration typically described as wind accretion of the Bondi-Hoyle-Lyttleton (BHL) type \citep{BHLreview}. Assuming the average ejecta speeds of the molecular and dusty outflow of 150\,\kms\ (Sect.\,\ref{sec-discussion-gas}), the accretion radius (a.k.a. Bondi-Hoyle radius) is 0.6\,au or 27 times the radius of the B star. In the absence of other effects, material with impact parameters smaller than this radius will be accreted in the wake of the flow through an unstable accretion column \citep{Ohsugi} (but see below). Particles with slightly larger impact parameters will be deflected (gravitationally focused) and may collide with each other in the wake of the flow, producing a turbulent region upstream. The region with a high velocity-dispersion and with altered composition of molecular gas seen in our ALMA maps near the companion is most naturally explained by this effect. Some particles are accelerated passing by the B star and this is perhaps why we observe the larger extent of the molecular remnant along the line connecting the two stars (see bottom right panel in Fig.\,\ref{fig-line-maps}).     

Although we invoke the BHL scenario here, we do not claim that there is significant ongoing accretion onto the B star. From CO observations we estimate that the H$_2$ density near the B star is about 10$^6$\,cm$^{-3}$ which gives accretion rates of about 10$^{-10}$\,M$_{\sun}$\,yr$^{-1}$. This is only an approximate value calculated under the multiple simplifying Bondi-Hoyle assumptions, but it implies a very low accretion luminosity of $\approx$0.006\,L$_{\sun}$, insignificant for the system energetics. 

Looking at the continuum maps, it is tempting to conclude that the curving tail of continuum emission is a continuous feature emanating from V838\,Mon and being sculpted by the gravity of the companion. The orbital motion of the companion of $\approx$4\,\kms\ is much slower than the ejecta, and therefore it does not cause any significant or systematic acceleration of the ejecta in directions other than the original flow motion. However, if our estimate of the orbital period of the system is roughly correct and the orbit is very close to the sky plane, at the moment of the eruption, $\sim$17\,yr ago, the B star had a location 6\degr\ off (with respect to the position of V838\,Mon). Assuming that the B star moves counterclockwise, the bright part of the extended dusty cloud was closer to the line connecting the stars 17\,yrs ago than it was in the ALMA epoch. If interaction with the B star formed the outer dusty cloud, it had to act on material that was lost from the inner system years prior to the eruption and one that had a chance to reach the B star. (The outer dust cloud, V838\,Mon, and the B star were all aligned near 1990, over a decade before the merger.) Although mass loss is expected in systems evolving into a merger years before the final coalescence \citep[e.g.][]{Tylenda2011,MacLeod2018a,Pejcha2016,Pejcha2016b}, assigning the outer dusty cloud to such a phenomenon is currently highly speculative. Also, it is unclear why the cloud is seen only in dust emission and not in gas tracers.

\subsubsection{Ionization, radiation pressure, and wind of the B star}
In 2005, the ejecta -- then approaching the B star -- was readily photo-ionized by radiation of the hot star \citep{TylendaEngulf}. That process very likely continued beyond 2005, but eventually the ejecta became too optically thick for the ionizing photons to penetrate deep into the outflow. The \ion{H}{II} region must currently have a small volume and in the optical is completely obscured by dust of the B component. 

The radiation of the star also provides significant pressure acting on the dust of the ejecta. Assuming a full collisional momentum coupling between dust and molecular gas, we estimate in Appendix\,\ref{appendix-radpress} that the acceleration caused by radiation of an unattenuated B3\,V star may even be ten times stronger than gravity. As also discussed in Appendix\,\ref{appendix-radpress}, however, sublimation of grains heated by stellar photons limits the expansive effect of radiation pressure, especially for small impact parameters. Once grains are destroyed, practically no momentum transfer from photons occurs (except for a weak effect in atomic lines). Overall, we expect that radiation pressure adds to the net scattering effect of material passing the B star and can be in part responsible for the relatively wide opening angle of the wake that is seen in the north-western part of the remnant in the continuum maps. In the case of an inefficient collisional coupling between dust and gas in the ejecta, the scattering effect of radiation pressure should be stronger for dust. 

The effect of radiation pressure acting on dust should produce a cone-like void behind the B star, very similar to that seen in SiO distribution. We modified the 3D model of the dust distribution presented in Sect.\,\ref{sec-3d-dust} by introducing a conical void whose apex is anchored at the B star, has an opening angle of $\sim$90\degr\ and the central axis pointing to V838\,Mon. The modification of the model has three main consequences. Firstly, the density of dust in the remaining part of the envelope surrounding the B star needs to be increased by about 30\% to match the flux of the ALMA observations. Secondly, although the B component still appears roughly circular, introducing the void shifts its centroid position $\sim$4\,mas closer to V838\,Mon. To reproduce the observations, the separation thus needs to be increased to 250\,au. Thirdly, the clearing around the B star exposes the external dust clouds to the unattenuated radiation of the B star and consequently 20\% less dust is needed to explain the observed flux levels of the extended cloud.

In addition to the radiation pressure effect, one could expect a repelling effect from a wind of the B star. Main-sequence early B-type stars have tenuous winds with mass-loss rates on the order of 10$^{-11}-10^{-10}$\,M$_{\sun}$\,yr$^{-1}$ and terminal velocities of 800--1330\,\kms\ \citep{Bwinds,nearMS}. As we show in Appendix\,\ref{appendix-wind}, such a wind will stop and shock the merger ejecta, but its dynamical effect on the shape of the flow wake is negligible. Collisions of wind and ejecta particles are expected to be quite energetic in V838\,Mon, but there are currently no observations that would directly identify the associated shock signatures in the environment of the red nova \citep{Antonini}. Nevertheless, such shocks may provide an extra local radiation source, heating dust in the zone where the two outflows collide. 

Given that nearly 17\,yr have passed since the eruption (in the ALMA epoch) and 15\,yr since the engulfment of the B star by the dusty ejecta, we would expect the companion to reappear for optical observations  --- most of the ejecta has passed the companion and gets geometrically thinner on our line of sight owing to the expansion. The reappearance has not yet happened \citep{Barsukova} either because the dusty ejecta has a large radial extent, which is unlikely, or part of the dusty material that obscured the B star became gravitationally bound. The latter would be surprising in face of the physical mechanisms discussed above which should clear of dust the immediate surroundings of the B star. 

\section{Summary}\label{sec-summary}
We report ALMA observations of the merger site of V838\,Mon 17\,yr after its outburst and at an unprecedented resolution corresponding to spatial scales of 150\,au. Continuum dust emission and emission in several molecular transitions were mapped and analyzed with radiative transfer calculations. The high-resolution maps show directly the complex interplay between material ejected during the red nova eruption and the remaining binary system. The inner remnant is dominated by a clumpy molecular wind similar to circumstellar envelopes of genuine red supergiants. Images of mm-wave continuum, supported by 3D radiative-transfer modeling, provide new evidence of a dusty elongated structure of about 95\,au in the intermediate neighborhood of the stellar remnant of V838\,Mon. The structure may be a dusty disk that stores the angular momentum of the binary that merged in 2002 and thereby is most interesting for studies of stellar mergers and common envelope events. Future observations of the disk require infrared interferometric techniques.  

The mm-wave continuum maps show for the first time the location of the B star, which has remained embedded in the dusty ejecta since 2005 and is not visible at shorter wavelengths. The orbit of the B star around V838\,Mon has a size of $\gtrsim$230\,au and, assuming a circular case, has a period of about 1000\,yr. These parameters make the V838\,Mon system most similar to Antares AB, where a wind of a red supergiant interacts with a B2.5\,V star at a similarly wide orbit. Instead of the slow (20\,\kms) wind of Antares A, in the case of the V838\,Mon system the hot star interacts with the much thicker and faster (150\,\kms) ejecta produced in the 2002 eruption of the red nova. This interaction will likely continue in the future, allowing us to better constrain the orbit of the B star and investigate whether it could have triggered the merger of the inner binary through the Kozai-Lidov mechanism \citep{Naoz}.

The combined ALMA data show complex interaction of the gravity and radiation of the hot star with the overflowing ejecta of V838\,Mon. The gravity pull accelerates the ejecta and creates a turbulent wake, which we identify in ALMA data as a region with an increased velocity dispersion and a changing chemical composition. Associated shocks create conditions where some species are enhanced (SO$_2$/H$_2$S) while others become depleted (SiO), presenting a unique laboratory of shock-excited circumstellar chemistry. The radiation pressure deflects part of the ejecta into a wide-angle cone and imposes differences in the distribution of gas and dust in the outer remnant. 

Comparing our ALMA data to results of spectroscopic monitoring of the object at shorter wavelengths, we find that external to the molecular remnant described here is fast atomic outflow, not seen at mm wavelengths, but which can be mapped with modern instruments at optical or IR wavelengths. Such observations would be necessary to assess the entire mass of the circumstellar material. Without accounting for the unknown contribution of the atomic phase, the estimate of the remnant's mass is currently $M_{\rm H2}$=0.1\,M$_{\sun}$. Our independent but uncertain estimate of the mass of dust only is 8.3$\cdot$10$^{-3}$\,M$_{\sun}$. The total mass and future characterization of the angular momentum stored in the inner disk will provide invaluable parameters for constraining the merger scenario of V838\,Mon and merger physics in general.

\begin{acknowledgements}
We thank Morgan MacLeod for discussions on possible interaction mechanisms in a wide binary of V838\,Mon. 
T.K. acknowledges funding from grant no 2018/30/E/ST9/00398 from the Polish National Science Center. R.T. and A.F. acknowledge a support from grant 2017/27/B/ST9/01128 financed by the Polish National Science Center. The research work at the Physical Research Laboratory is funded by the Department of Space, Government of India. M.S. acknowledges a support by the National Science Center, Poland, under grant 2016/21/B/ST09/01626.

This paper makes use of the following ALMA data: ADS/JAO.ALMA\#2018.1.00336.S. ALMA is a partnership of ESO (representing its member states), NSF (USA) and NINS (Japan), together with NRC (Canada), MOST and ASIAA (Taiwan), and KASI (Republic of Korea), in cooperation with the Republic of Chile. The Joint ALMA Observatory is operated by ESO, AUI/NRAO and NAOJ. The National Radio Astronomy Observatory is a facility of the National Science Foundation operated under cooperative agreement by Associated Universities, Inc.

Some observations reported in this paper were obtained with the Southern African Large Telescope (SALT). Polish participation in SALT is funded by grant No. MNiSW DIR/WK/2016/07. Based on observations collected at the European Organization for Astronomical Research in the Southern Hemisphere under ESO program 088.D-0112(A). 

This research is based in part on data collected at Subaru Telescope, which is operated by the National Astronomical Observatory of Japan. We are honored and grateful for the opportunity of observing the Universe from Maunakea, which has the cultural, historical and natural significance in Hawaii.

Based in part on observations made with the NASA/DLR Stratospheric Observatory for Infrared Astronomy (SOFIA). SOFIA is jointly operated by the Universities Space Research Association, Inc. (USRA), under NASA contract NNA17BF53C, and the Deutsches SOFIA Institut (DSI) under DLR contract 50 OK 0901 to the University of Stuttgart. 

\end{acknowledgements}

\begin{appendix}
\section{Optical spectra}\label{appendix-optical} 
We used the X-SHOOTER spectrograph on the Very Large Telescope (VLT) to obtain spectra of V838\,Mon. The observations were executed on 6 January 2012 in nodding mode as program 088.D-0112(A). The spectra cover the range 0.3--2.4\,$\mu$m. Slit widths of 1.3, 0.7, and 0.4 arcsec were used for UV, VIS and NIR arms of the spectrograph \citep{xshooter}, respectively.  This resulted in a resolution of about 11\,500 in the VIS and NIR arms, and of 4100 in the NIR part. Additionally, spectra with a wide slit of 5\arcsec\ were secured for a better absolute flux calibration. All spectra were reduced with the most recent version of the {\it esorex} pipeline (source kit 3.5.0 from May 2020) in the ESO Reflex environment. Three exposures in each arm were taken and combined. Spectra were corrected for telluric absorption using the standalone {\it Molecfit} program \citep{molecfit1,molecfit2}. The correction was calibrated on carefully selected bands in the spectra of the science object, avoiding water bands intrinsic to the star. A correction to the Solar System barycenter was applied and spectra were corrected for the interstellar extinction with $E_{B-V}$ of 0.85\,mag and $R_V$=3.1.   

An optical spectrum in the range 370--890\,nm was acquired at the South African Large Telescope (SALT) telescope under a project number 2020-2-DDT-001. The observations were taken on 23 December 2020 with the High Resolution Spectrograph \citep{hrs} in its high-resolution mode and with $R\!\approx$40\,000. Data were automatically reduced with the default instrument pipeline\footnote{\url{http://pysalt.salt.ac.za/}} \citep{SALTpipeline}. Spectra from five exposures were averaged. Shortward of 4500\,\AA, the spectrum is underexposed with a signal-to-noise ratio of a few. The SALT spectra are not calibrated in absolute flux units. No correction for telluric spectrum could be applied. We used spectra corrected to the Solar System barycenter.

\section{SED data sources}\label{appendinx-a}
The data used in our SED reconstruction are listed in Table\,\ref{tab-sed}. 

The $UBVR_CI_C$ photometry was taken from \citet{Barsukova} through the website of V.~Goranskij\footnote{\url{http://www.vgoranskij.net/v838mon.ne3}}. We averaged measurements within about 0.5\,yr from the epoch of ALMA observations. Three standard deviations related to source variability in that period added to the photometric uncertainties given in \citet{Barsukova} serve as our effective measurement error. The broadband magnitudes were transformed to flux units using generic zero points of the Johnson and Cousins photometric systems for Vega.  

Near-infrared photometric observations were carried out on UT 16.72 February 2019 at the 1.2\,m telescope of the Mount Abu Infrared Observatory in India and using the Near-Infrared Camera/Spectrometer (NICS) equipped with a 1024$\times$1024 HgCdTe Hawaii\,1 array. The camera has an unvignetted 8\arcmin$\times$8\arcmin\ field of view and uses $JHK_s$ filters that conform to Maunakea Observatories (MKO) specifications. Frames in each filter were obtained in five dithered positions, typically offset by 40\arcsec, and with multiple frames (typically five) obtained in each dithered position. The total exposure time in the individual filter was 8\,s. Standard procedures of dark and flat field corrections were applied. The corrected science frames were added to produce an average frame on which photometry was performed. Aperture photometry was derived using routines in IRAF, with the 2MASS field stars used for photometric calibration. The measured magnitudes $J$=6.89($\pm$0.03), $H$=5.86($\pm$0.03), and $K$=5.08($\pm$0.06) were converted to flux units using 2MASS zero points. The errors given here have a meaning of 3$\sigma$.  

We extracted data observed with the Wide-field Infrared Survey Explorer (WISE) using the WISE point source catalog \citep{WISE}. They represent the full period of WISE observations of V838\,Mon, that is, from April to October 2010. Catalog magnitudes were converted to fluxes using standard zero points and with no color corrections. Uncertainties have a meaning of 3$\sigma$. 

On UT 29 January 2018, V838\,Mon was observed with COMICS \citep{kataza00,okamoto03} on the Subaru 8.2\,m telescope in imaging mode with the $N12.4$ filter (12.4\,$\mu$m central wavelength and a width of 1.2\,$\mu$m). COMICS in imaging mode feeds a 320 $\times$ 240 pixel Si:As array whose 0\farcs13\,pixel$^{-1}$ plate scale affords a field of view of 41\farcs6$\times$ 31.2\farcs. Observations were chopped only (keeping all beams on chip and lowering overhead) to maximize efficiency. The chopping throw used was 10\arcsec\ along a position angle of $\approx$0$^{\circ}$. Chop pairs are differenced to remove the rapidly fluctuating background signal, then combined to yield the final reduced image. The flux calibration standard star HR\,2639 (K5\,III) was observed immediately before V838\,Mon, and the adopted flux density
for it in the $N12.4$ band was determined from the WISE $W3$ measurement scaled to 12.4\,$\mu$m.

Far-infrared imaging photometry was obtained for V838\,Mon on UT 24 October 2017 with HAWK+ \citep{harper18} onboard the Stratospheric Observatory for Infrared Astronomy (SOFIA). Images were obtained in total intensity mode in Bands A and C with Lissajous scan imaging. Data were reduced with {\sf CRUSH} within the HAWC+ facility data reduction pipeline version 1.3.0. Level~3 pipeline-reduced and calibrated data products are obtained from the SOFIA science archive. Aperture photometry is performed directly on the flux-calibrated images with an aperture size sufficient to encircle all flux from V838\,Mon (for Band C the source image appears partially elongated along one direction, most likely due to an instrument or aircraft artifact); it is not clear if an aperture correction is necessary to compensate. Flux uncertainties are estimated from the standard deviation of ten photometric extractions with the same aperture made at random positions in the blank field around V838\,Mon.


For reference, we add to the SED in Fig.\,\ref{fig-sed} MIR and FIR measurements from 2007. These are data from {\it Spitzer} and Gemini telescopes. They were reported by \citet{WisniewskiPhoto}.

For the comparison with the simulated SED, all data were corrected for interstellar extinction with $E_{B_V}$=0.85\,mag and $R_V$=3.1, following  \citet{TylendaProgenitor}.

\begin{table}
    \centering
    \caption{Flux measurements of V838\,Mon.}\label{tab-sed}
    \begin{tabular}{ccccc}
\hline
Filter & $\lambda_{\rm eff}$ &Flux &Relative &  Ref.\\
       &        ($\mu$m)     &(Jy) & 3$\sigma$ error & \\
\hline       
$U$	       &0.37	&0.07e-3	&0.7& 1 \\
$B$	       &0.44	&1.80e-3	&0.4& 1 \\
$V$	       &0.55	&1.45e-2	&0.3& 1 \\
$R_C$      &0.65	&9.06e-2	&0.2& 1 \\
$I_C$      &0.79	&0.5441	    &0.3& 1 \\
$J$	       &1.25	&2.796	    &0.03 & 2 \\
$H$	       &1.64	&4.638	    &0.03 & 2 \\
$K_S$      &2.20    &6.193	    &0.05 & 2 \\
$W1$       &3.35	&4.2	    &0.7  & 3 \\
$W2$       &4.60	&5.6	    &0.5  & 3 \\
$W3$       &11.56	&35.8	    &0.12  & 3 \\
$N12.4$&12.40	&16.1	    &2.4  & 2 \\
$W4$       &22.09	&40.4	    &0.003 & 3 \\
HAWK+ A	   &53.09	&11.5	    &3.6  & 2 \\
HAWK+ C	   &88.67	&4.1	    &1.5  & 2 \\
ALMA B6    &1333.87 &2.3e-3	    &0.5   & 2 \\
\hline
    \end{tabular}
    \tablebib{
     (1)~\citet{Barsukova}; (2) this study; (3)~\citet{WISE}.
             }
\end{table}

\section{Channel maps of SiO emission}\label{appendix-SiO}
\begin{figure*}
    \centering
    \sidecaption
    \includegraphics[width=0.8\textwidth, trim=0 0 0 0]{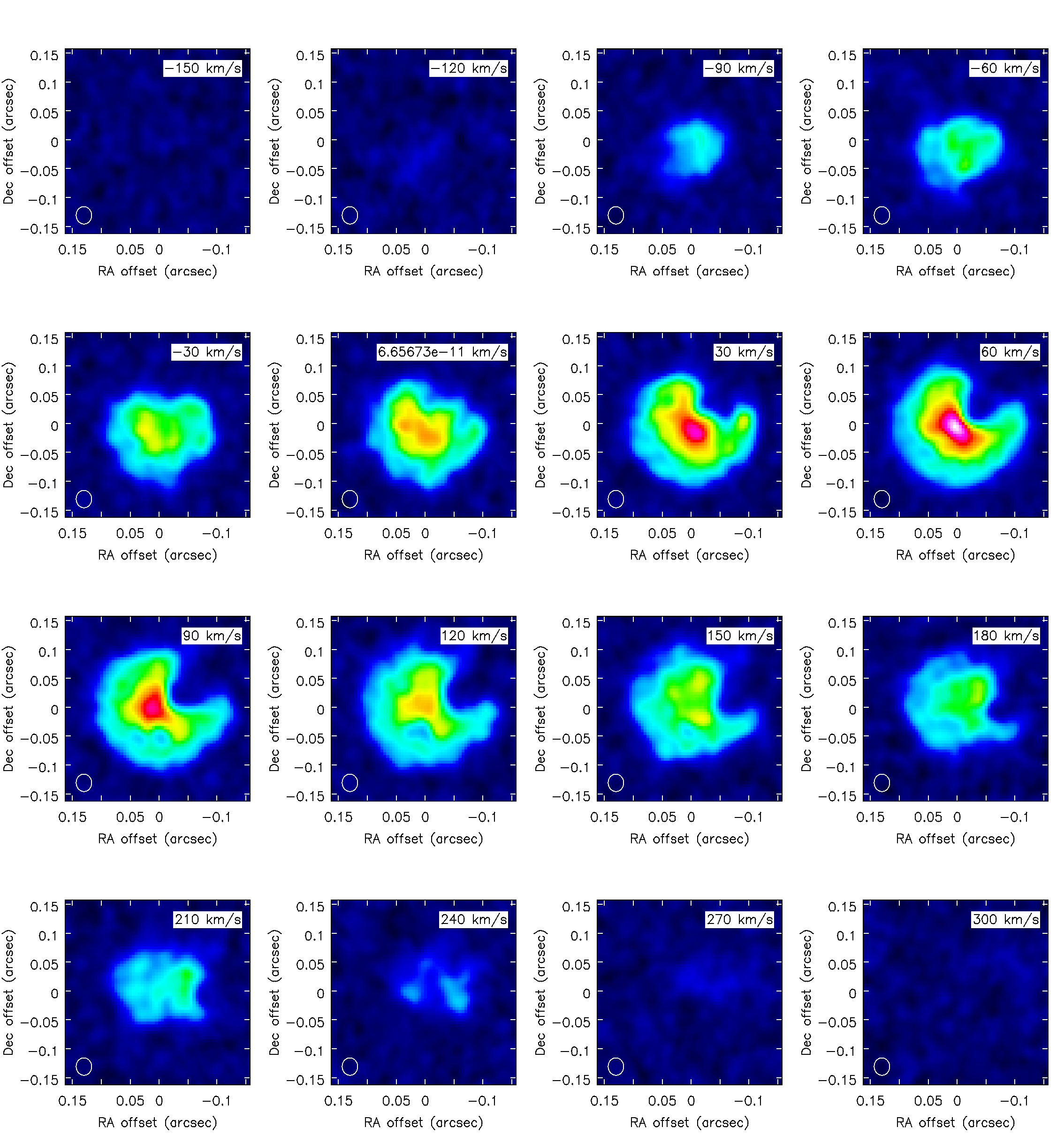}
    \includegraphics[trim=1300 200 100 0,clip, width=0.15\textwidth]{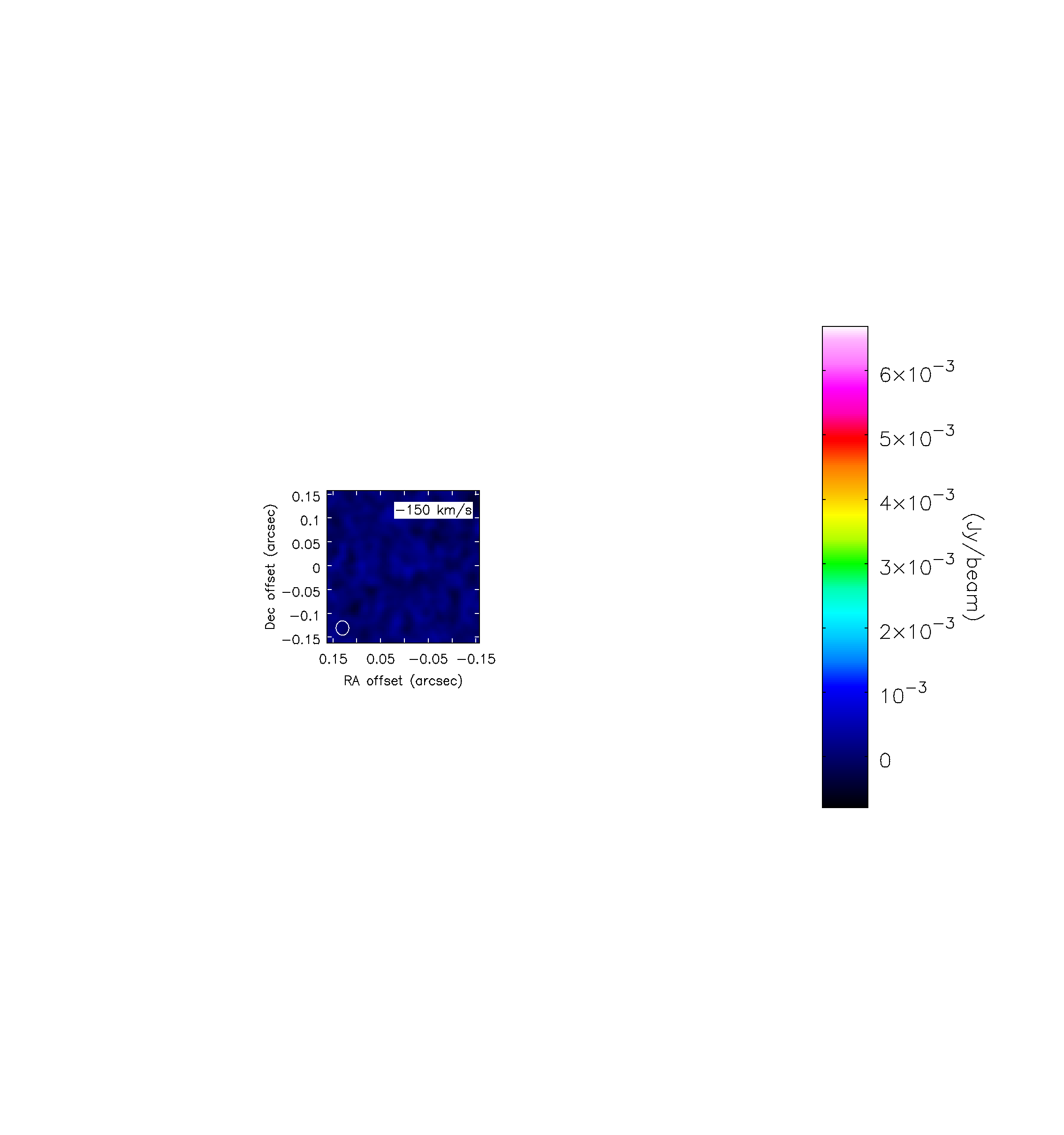}
    \caption{Channel maps of SiO emission at Briggs weighting with the robust parameter of 0.5.}\label{fig-channelmaps-SiO}
\end{figure*}

Figure\,\ref{fig-channelmaps-SiO} presents maps of SiO emission at selected velocities. These maps were used to construct the shape of the SiO region shown in Fig.\,\ref{fig-SiOshape} and discussed in Sect.\,\ref{sec-3d-mol}.

\section{Effects of radiation pressure of the B star on the ejecta}\label{appendix-radpress}

When approaching the B3V star, the flow from V838\,Mon is affected by the gravity of the star and the radiation pressure. The effective acceleration, $a_{\rm eff}$, of the matter at a distance $r$ from the star of a mass $M_s$ and a luminosity $L_s$ is thus
$$ a_{\rm eff} = g + a_{rad}$$
where
$$g = - G M_s/r^2$$
and
$$a_{rad} = (L_s/(4\pi r^2 c))~(\pi a^2 Q_{pr})/((4/3)\pi a^3\rho_d/R_{dg}).$$
Here, $a$ is the dust grain radius, $Q_{pr}$ is the radiation pressure efficiency factor, $\rho_d$ is the density of the grain material, $R_{dg}$ denotes the dust-to-gas mass ratio, and $G$ is the gravitational constant.

\begin{figure}
\centering
\includegraphics[width=0.8\columnwidth]{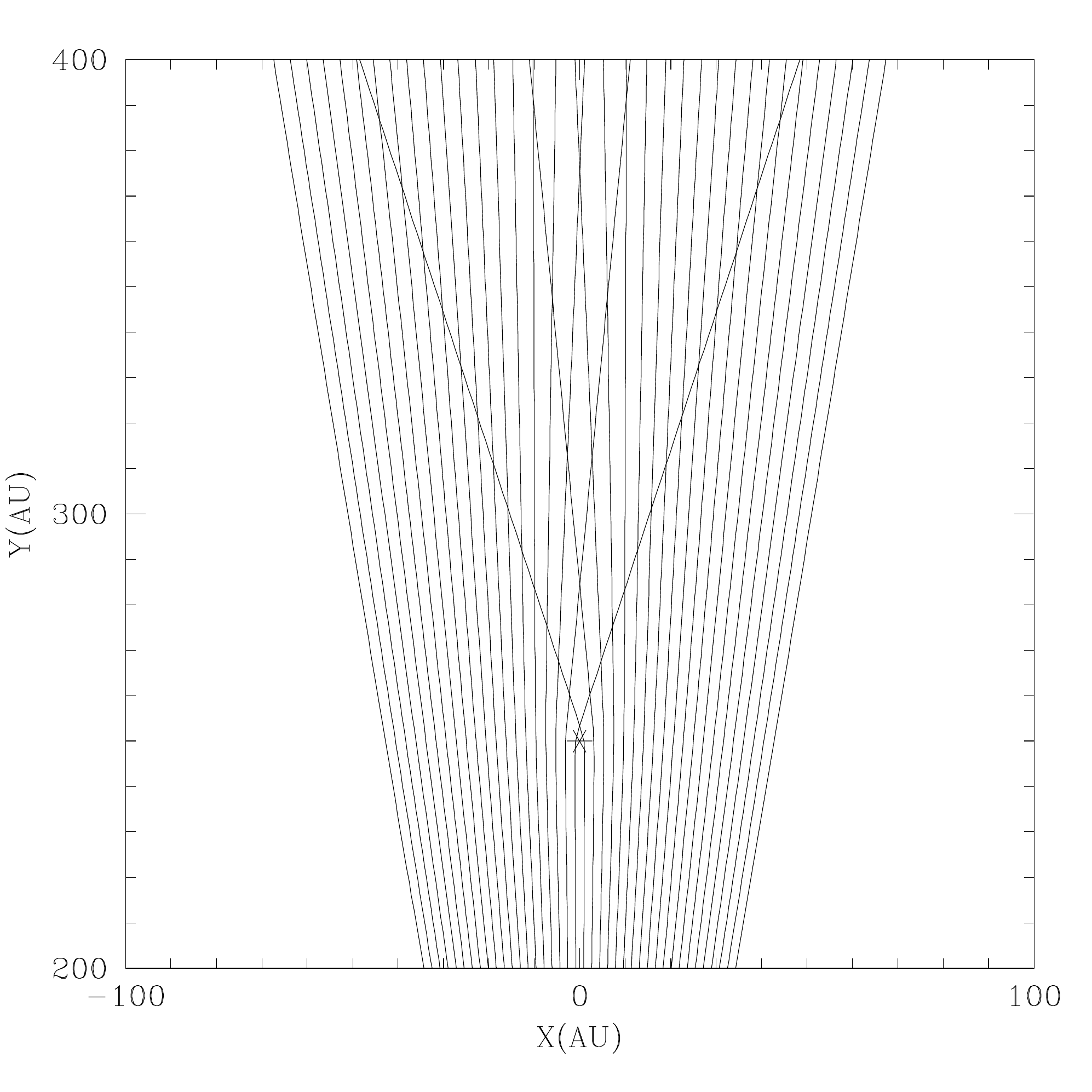}
\caption{Particle trajectories ($v_0 = 200$\,\kms)  in the vicinity of the B3\,V companion in the case of $a_{\rm eff} = g$ (pure gravity).}\label{fig-tyl-1}
\end{figure}

Both, $g$ and $a_{rad}$ have the same dependence on the distance from the star ($r^{-2}$), so their ratio is independent of the distance. Therefore, it is convenient to rewrite the above formula as
$$a_{\rm eff} = g(1+a_{rad}/g)
  = g(1- (3/16\pi) L_s/(G M_s c)~(Q_{pr}R_{dg}/(a\rho_d)).$$
Substituting values of the physical constants and expressing the stellar luminosity, $l_s$, and mass, $m_s$, in solar units,  one obtains
$$a_{\rm eff} = g(1 - 5.74\,10^{-5} (l_s/m_s)(Q_{pr} R_{dg}/(a \rho_d)).$$
We adopt typical parameters of a B3\,V star, i.e. $m_s = 7.0$, $l_s = 1800$, as well as T$_{\rm eff} \simeq 18\,000$\,K and R$_s \simeq 5.0$\,R$_s$. We also adopt "astronomical silicate" dust with $\rho_d \simeq 2.0$\,g\,cm$^{-3}$, $R_{dg} = 0.01$, and $a = 0.1\mu m$. For the above effective temperature and grain radius, $Q_{pr} \simeq 1.4$ (from the B. Draine's web page\footnote{\url{https://www.astro.princeton.edu/~draine/dust/dust.diel.html}}). Substituting these values, yields
$$a_{\rm eff} \simeq g(1 - 10.5) = -9.5 g.$$
Thus, in the absence of other effects, the flow around the B3\,V companion is dominated by the stellar radiation pressure on the dust grains, rather than the stellar gravitational field. To illustrate the expected effects on the flow in the vicinity of the B star, we integrated the standard (ballistic) equation of motion for particles ejected by V838\,Mon with a given velocity in directions towards the companion. The acceleration in the above form is the only factor governing the motion of the particles. No (hydrodynamic) effects of possible crossing of the particle trajectories were taken into account. Figure\,\ref{fig-tyl-1} shows the case of pure gravity ($a_{rad}$ = 0). V838\,Mon is at the coordinate origin ($x$=0 and $y$=0), B3\,V star is at $x$=0 and $y$=250\,au (shown by an asterisk in the figure). Particles are injected at (0,0) with a velocity of 200\,\kms\ and at different (small) angles from the $x$ axis (i.e. a line joining V838\,Mon and B3\,V). Due to the gravity of the companion, the trajectories are bent toward the axis of symmetry ($x$=0) and those passing close to the star intersect behind the star. In reality, we expect that a turbulent wake is produced in the flow behind the perturbing star. 

\begin{figure}
\centering
\includegraphics[width=0.8\columnwidth]{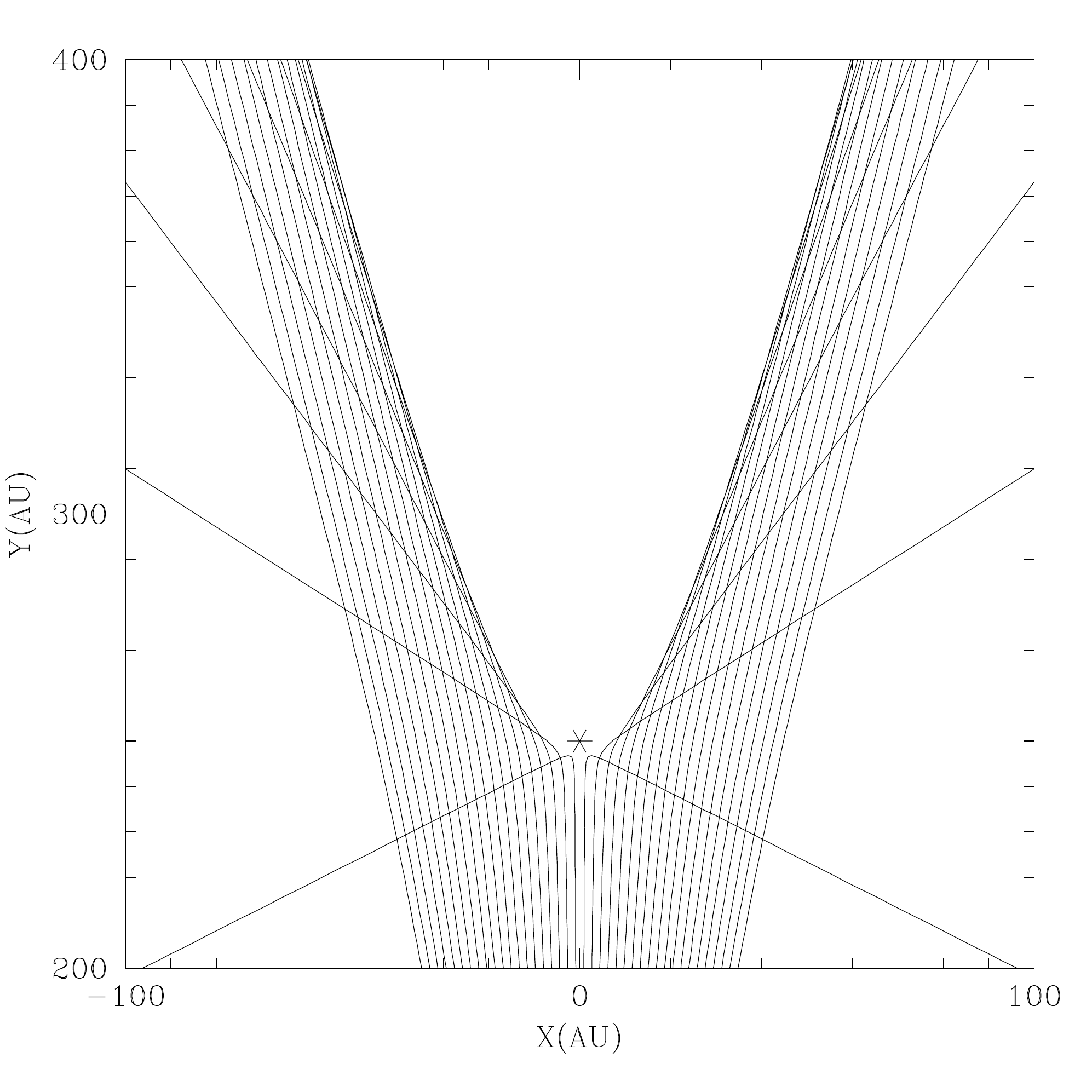}
\caption{The same as in Fig.\,\ref{fig-tyl-1} but for $a_{\rm eff} = -10g$.}\label{fig-tyl-2}
\end{figure}

Figure\,\ref{fig-tyl-2} presents the case when $a_{\rm eff} = -10g$. Other parameters remain the same as above. The radiation pressure slows down and rejects the particles approaching the star too closely. This creates an empty cone in the immediate vicinity of and behind the B star. The effect is stronger when the flow is slower. The trajectories intersect, especially in the vicinity of the star. In reality, this should result in an increase of the density of the flow and disordered (turbulent) motions, especially near the edge of the cone. Since the velocities involved are highly supersonic, a creation of shock fronts is also likely.

\begin{figure}
\centering
\includegraphics[width=0.8\columnwidth]{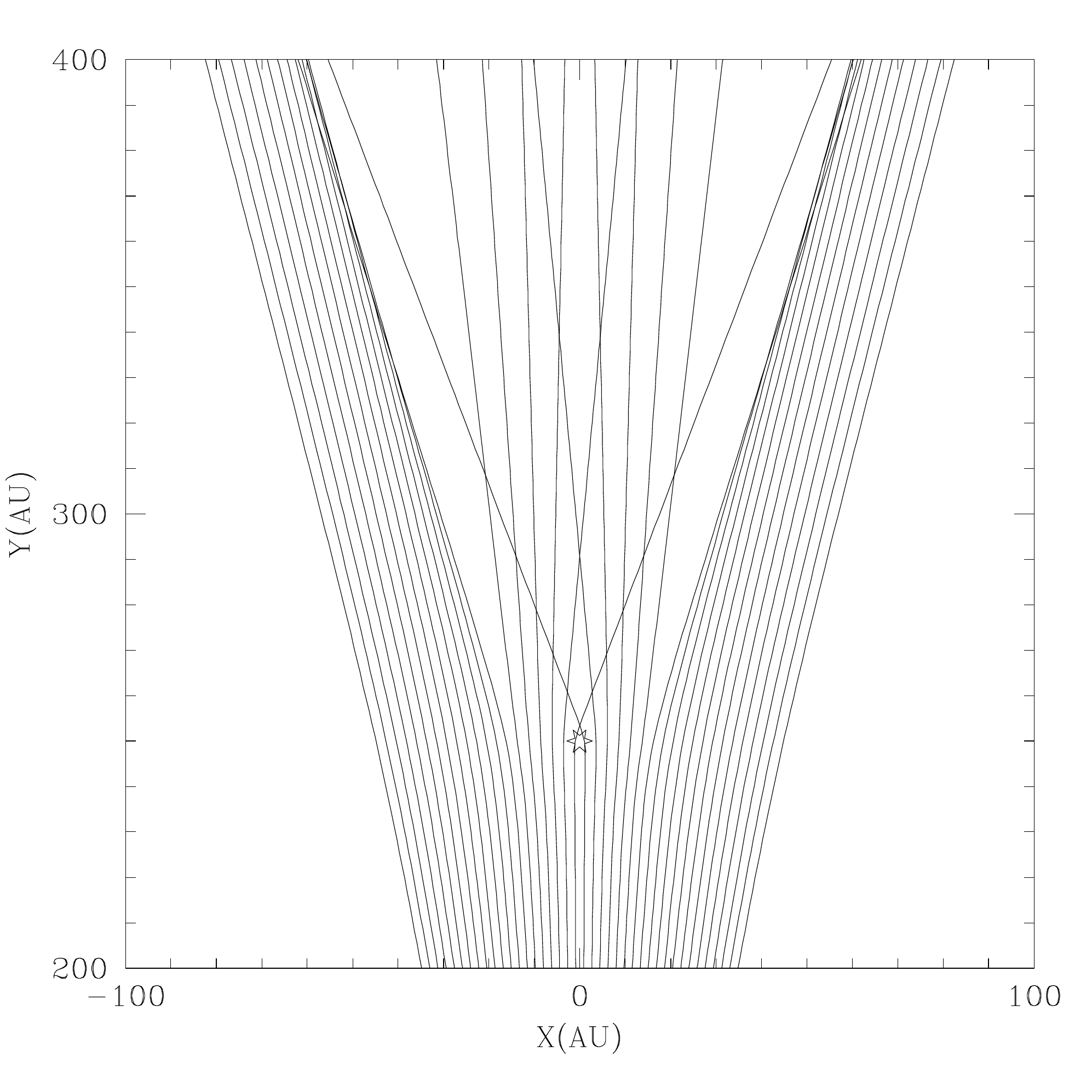}
\caption{The same as in Fig.\,\ref{fig-tyl-2} but with dust evaporation at $r < r_d$.}\label{fig-tyl-3}
\end{figure}

The above considerations do not take into account dust sublimation that is expected to take place close to the star. If dust is destroyed at a sublimation temperature, $T_d$, then it takes place at a distance, $r_d$, which can be estimated from
$$(4\pi R_s^2 \sigma T_s^4)/(4\pi r_d^2)~\pi a^2 Q_{\rm abs}(T_s) =
  4\pi a^2 \sigma T_d^4 Q_{\rm em}(T_d).$$
Here $R_s$ is the radius of the star whereas $Q_{\rm abs}$ and $Q_{\rm em}$ are the Planck-averaged dust absorption and emission coefficients, respectively. The above equation can be rewritten as
$$(R_s/r_d)^2 = 4 Q_{\rm em}/Q_{\rm abs}~(T_d/T_s)^4,$$
or
$$r_d/R_s = (Q_{\rm abs}/4Q_{\rm em})^{1/2} (T_s/T_d)^2.$$
Assuming that the silicate grains sublimate at $T_d \simeq 1250$\,K \citep{Tsub} and taking the respective $Q$ values from the website of B. Draine ($Q_{\rm abs}(18\,000$\,K)$/Q_{\rm em}(1250$\,K)$ \simeq $ 31.2), we finally obtain 
$$r_d \simeq 580\,R_s= 13\,{\rm au}.$$
(In our 3D model introduced in Sect.\,\ref{sec-3d-dust}, which was constructed with slightly different grains and stellar parameters, the clearing around the B star is 10\,au, consistent with the calculation here.) We next assume that if the matter is able to penetrate inside $r_d$, then there is no dust, and consequently $a_{rad}=0$ and $a_{\rm eff}=g$. We further assume that dust is not formed again when the matter leaves the region where $r < r_d$. Figure\,\ref{fig-tyl-3} shows the results of including the sublimation effect. Particles ejected with larger angles from the $x$ axis do not approach the star closer than $r_d$, so their trajectories are the same as in Fig.\,\ref{fig-tyl-2}. On the other hand, particles ejected at small angles are able to penetrate closer than $r_d$ and their trajectories are similar (but not the same) as in Fig.\,\ref{fig-tyl-2}. Going back to the realistic case of the V838\,Mon ejecta and its companion, we can expect that with dust sublimation, the B3\,V companion can form both an extended cone of compressed matter and a turbulent wake. 

\begin{figure}
\centering
\includegraphics[width=0.8\columnwidth]{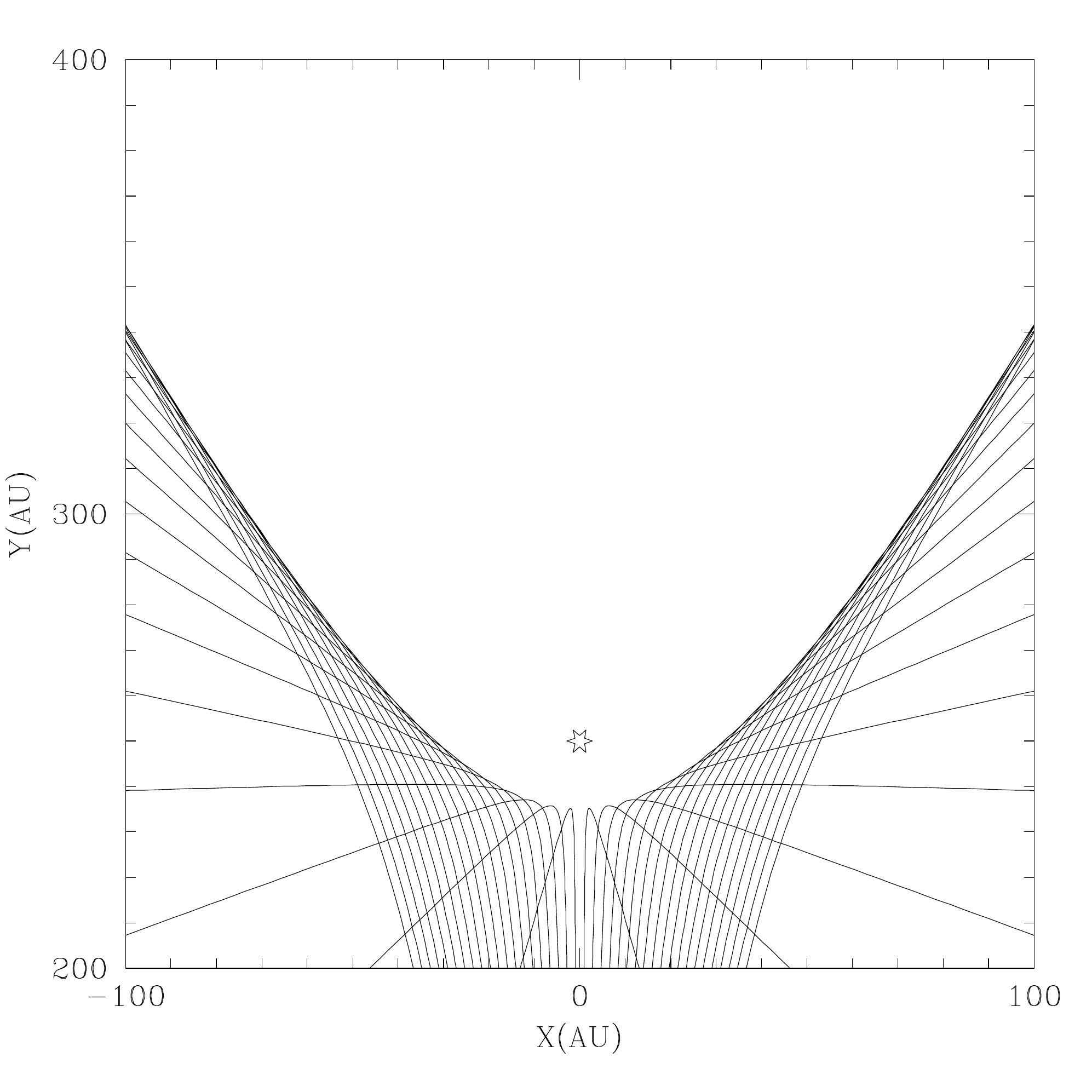}
\caption{The same as Fig.\,\ref{fig-tyl-3}  but for $v_0 = 95$\,km\,s$^{-1}$.}\label{fig-tyl-4}
\end{figure}

Note that the results are sensitive to the velocity of the flow. For instance, for $v_0 \simeq$ 95\,\kms, the matter is nowhere able to penetrate inside $r_d$ and no wake is formed, as illustrated in Fig.\,\ref{fig-tyl-4}.

\section{Shocks induced by the wind of the B star}\label{appendix-wind}

As noted above, the stellar radiation pressure potential has the same  dependence on the distance from the star as the gravitational potential (with the opposite sign).  Therefore, by analogy to the formula for the free-fall velocity in the gravitational field, the stopping radius, $r_{stop}$, at which the inflowing dusty matter is stopped in the stellar radiation field, can be derived from
$$v_0 = (2 f G M_s/r_{\rm stop})^{1/2}, $$
where $f = a_{\rm eff}/g$. Thus, 
$$r_{\rm stop} = 2 f G M_s/v_0^2. $$
The distance from the star, $r_{\rm sh}$, where the stellar wind stops and shocks the inflowing matter can be estimated from a standard condition of momentum flux equilibrium, that is
$$ N_0 v_0^2 = N_w v_w^2, $$
where the subscript ``0'' refers to the flow parameters, while "$w$" -- to those of the stellar wind. Thus,
$$ N_0 v_0^2 = v_w \dot{M}_w/(4\pi r_{\rm sh}^2 m_p), $$
where $\dot{M}_w$ is the wind mass-loss rate and $m_p$ denotes the particle mass. Finally,
$$ r_{\rm sh} = (\dot{M}_w v_w/(4\pi m_p N_0 v_0^2))^{1/2}. $$

Let us apply the above formulae to the parameters expected for the B3\,V companion and the observed outflow from V838\,Mon in the vicinity of the companion, with $v_0$=75--250\,\kms, $N_0$=10$^{6-7}$\,cm$^{-3}$, $\dot{M}_w = (1-4)\times10^{-11} M_\odot$\,yr$^{-1}$, and $v_w$= 800--1400\kms. The calculated radii are given in Table\,\ref{tab-wind}.
\begin{table}
    \centering
    \caption{Radii where ejecta is stopped and shocked by the wind of the B star.}\label{tab-wind}
    \begin{tabular}{ccccccc}
\hline
$v_0$&$N_0$&$\dot{M}_w$&$v_w$&$r_{\rm stop}$&$r_{\rm sh}$&$r_d$\\
\kms&cm$^{-3}$&M$_{\sun}$\,yr$^{-1}$&\kms&au&au&au\\
\hline       
250&$10^7$&1\,10$^{-11}$&~800&~~2.3&0.04&14.5\\
250&$10^7$&4\,10$^{-11}$&1400&~~2.3&0.11&14.5\\
~75&$10^6$&1\,10$^{-11}$&~800&25.3& 0.44&14.5\\
~75&$10^6$&4\,10$^{-11}$&1400&25.3& 1.15&14.5\\
\hline \end{tabular}
\tablefoot{For comparison, $r$(B3\,V) = 5.0 R$_\odot$ = 0.02\,au}.
\end{table}
They can be summarized as follows. About 5\,yr after the V838\,Mon eruption, the fastest and probably the densest ejecta reached the B3\,V companion. Dust was evaporated well before the matter could have been stopped by radiation
pressure on dust grains ($r_d \gg r_{\rm stop}$). As a result, the matter could have approached the companion until it had been shocked and stopped by the stellar wind a few stellar radii above the stellar surface. Probably no matter was then accreted on the companion. Instead, a hot shocked region was then formed above the photosphere. 

At the ALMA epoch and at present, the matter that flows in the vicinity of the companion is much slower and probably less dense. Therefore, it is stopped by the radiation pressure about 1000 stellar radii above the stellar photosphere, 
well before the dust grains can evaporate. No matter is accreted on the companion. The wind collides with the inflowing matter at $r_{\rm stop}$ but its dynamical influence is negligible. A rarefied hot region is probably formed near $r_{\rm stop}$.
\end{appendix}

\bibliographystyle{aa}
\bibliography{bib.bib}

\end{document}